\documentclass{article}

\usepackage[english]{babel}

\usepackage[letterpaper,top=2cm,bottom=2cm,left=3cm,right=3cm,marginparwidth=1.75cm]{geometry}

\usepackage{amsmath}
\usepackage{graphicx}
\usepackage[colorlinks=true, allcolors=blue]{hyperref}

\usepackage{xcolor}
\usepackage{comment}
\usepackage{hyperref}
\usepackage{caption}
\usepackage{amsmath}
\usepackage{bbold}
\usepackage{etoolbox}
\usepackage{adjustbox}
\usepackage{makecell}
\usepackage{subcaption}
\usepackage{textcomp}
\usepackage[symbol]{footmisc}
\usepackage{booktabs}

\usepackage{algorithm}
\usepackage{algorithmic}
 
\usepackage[sort&compress,numbers]{natbib}
\providecommand{\keywords}[1]{\textbf{Keywords:} #1}

\title{Physics Informed Neural Networks for Modeling of 3D Flow-Thermal Problems with Sparse Domain Data}

\author{\stepcounter{footnote}Saakaar Bhatnagar$^{a}$\thanks{Corresponding Author} \\ {\small \textit{sbhatnagar@altair.com}} \and Andrew Comerford$^{a}$ \\ { \small \textit{acomerford@altair.com}} \and  Araz Banaeizadeh$^{a}$ \\ {\small \textit{araz@altair.com}}}

\date{$^{a}${\small Altair Engineering Inc., 100 Mathilda Place, Sunnyvale, CA, USA}} %leave blank

\begin{document}

\maketitle

\keywords{Physics Informed Neural Networks; Navier-Stokes Equations; Surrogate Modeling; Design Optimization} \\

\begin{abstract}
  \centering Successfully training Physics Informed Neural Networks (PINNs) for highly nonlinear PDEs on complex 3D domains remains a challenging task. In this paper, PINNs are employed to solve the 3D incompressible Navier-Stokes (NS) equations at moderate to high Reynolds numbers for complex geometries. The presented method utilizes very sparsely distributed solution data in the domain. A detailed investigation on the effect of the amount of supplied data and the PDE-based regularizers is presented. Additionally, a hybrid data-PINNs approach is used to generate a surrogate model of a realistic flow-thermal electronics design problem. This surrogate model provides near real-time sampling and was found to outperform standard data-driven neural networks when tested on unseen query points. The findings of the paper show how PINNs can be effective when used in conjunction with sparse data for solving 3D nonlinear PDEs or for surrogate modeling of design spaces governed by them.
\end{abstract}

\section{Introduction}
\label{sec:intro}

Over the last few years, there has been significant growth in the popularity of machine learning algorithms to solve partial differential equations (PDE) or assist PDE solvers, such as computational fluid dynamics (CFD) solvers \citep{brunton_ML_fluids,DL_cfd_aid}. A particular application where CFD solvers struggle, due to the computational cost, is iterative design optimization. This is the process of continually updating a design (e.g. an electronics assembly layout) and computing the solution (e.g. flow or thermal fields) to optimize the performance (e.g. constrain the temperatures or reduce the pressure drop).  The challenge for CFD is the input-output relationship is one-to-one. Therefore, any changes to the input vector (e.g. geometric variations) need to be re-simulated, leading to high costs when iterating on different design scenarios \cite{Knight2001}. Overall high-fidelity iterative design requires a prohibitive level of resources, both computationally and monetarily, and often leads to a sub-optimal outcome. The attraction of Machine Learning (ML) algorithms in these scenarios is the ability to rapidly find solutions for such problem setups that are challenging in conventional CFD, such as large design space explorations \cite{bayes_opt_airfoil}, turbulence model closure \cite{turb_model_closure} or solving incomplete/ill-posed problems \cite{heat_transfer_PINNs}.

 Conventional ML algorithms usually require large amounts of data to train. This represents a challenge when using ML in engineering applications such as CFD, since experimental data can be difficult and expensive to obtain and may suffer from measurement noise. Furthermore, in many engineering experiments, field data such as temperature and velocity fields can sometimes only be captured at specific locations, and it is difficult to get full field solution results from physical experiments. Research has turned to using simulation data for training ML models, but the computational cost of generating large amounts of data to train models is a major bottleneck.

Physics Informed Neural Networks (PINNs) \cite{PINN_orig} represent an advance in scientific machine learning that has the potential to solve many of the aforementioned issues. By adding the physics that governs the problem into the loss function, and optimizing the loss, it is possible to have the network learn the solution of the problem represented by that equation in a data-free manner. PINNs can be used in cases where sporadic experimental field data is available \cite{cai2021flow,jagtap2022physics} to calculate the rest of the field variable and can be used to solve problems with incomplete or missing physics \cite{chen2020physics,ISHITSUKA2023120855}. 

Another application area, in which PINNs could be very beneficial is machine learning-based surrogate modeling. Although a relatively new field, several ML architectures and methods have been utilized in the literature. These include: Proper Orthogonal Decomposition (POD) \cite{pod}, Gappy POD \cite{gappy_pod} and Manifold Learning \cite{manifold_learning}. More recently, increased attention has been given to statistical methods like Gaussian processes and neural networks that incorporate Machine Learning (ML) to create surrogate models. \citet{Bhatnagar2019} used a CNN architecture to predict aerodynamic flow fields over airfoils and created a surrogate model that generalized between flow conditions and airfoil geometries. \citet{guo_cnn} also used a Convolutional Neural Network (CNN) architecture to predict steady flows over automotive vehicles. \citet{Lee_2019} used Generative Adversarial Networks (GANs) coupled with physical laws to predict unsteady flow around a cylinder, demonstrating the benefits of using embedded physics. \citet{hidden_physics_GP} use Gaussian processes to model and identify several complex PDEs.

Several of the aforementioned studies used purely data-driven models and required the creation of large amounts of training data to generate accurate and generalizable models. PINNs have the capability to greatly reduce these data generation costs, and it has been shown that training surrogates using the physics embedded in the loss function greatly improves predictive accuracy, across a wide range of applications \cite{Lee_2019,pinn_nuclear,kim2023modeling, wang2021physics}.

However, there is currently a lack of research articles applying PINNs to 3-dimensional (3D) problems, particularly for highly nonlinear PDEs like the Navier-Stokes equations. These problems are challenging for PINNs due to a variety of reasons that are discussed later in this paper. Yet, these problems are the most lucrative to solve, as most industrial applications of CFD are done in 3D. This paper provides results that aim to address this gap, by solving several problems with realistic physical parameters, over complex geometries in a data-assisted manner, using very sparse domain data. Further, this paper solves a realistic flow-thermal design optimization problem using a hybrid data-PINN surrogate model and shows how PINN models outperform standard data-driven neural network (NN) surrogates for every test point queried in the Design of experiments (DoE) space for the surrogate modeling problem.

The paper is divided as follows; Section \ref{sec:PINNs} introduces PINNs in more detail and discusses some of the technical challenges with training PINNs. Section \ref{sec:impfeatures} outlines some of the important features the authors incorporate in the creation and training of PINNs to enable accurate and fast convergence. Section \ref{sec:exp_and_results} demonstrates several problems solved using PINNs, and showcases a design optimization problem using PINN-based surrogates. Section \ref{sec:conclusions} discusses how the work shown in this paper can be improved upon.

\section{Physics Informed Neural Networks (PINNs) }
\label{sec:PINNs}

\subsection{Setting up a PINN Training}
Physics-informed neural networks (PINNs) leverage automatic differentiation to obtain an analytical representation of an output variable and its derivatives, given a parametrization using the trainable weights of the network. By employing the underlying static graph, it is possible to construct the differential equations that govern physical phenomena.

A PDE problem in the general form reads:

\begin{equation}
\label{equation:PDE_main}
  \mathcal{N}_{\textbf{x}}[u]=0, \textbf{x} \in \Omega,
\end{equation}

\begin{equation}
\label{equation:PDE_BC}
\Phi(u(\textbf{x}))= \textbf{g}(\textbf{x}), \textbf{x} \in \partial \Omega 
\end{equation}

where \(\Phi\) can be the identity operator (Dirichlet B.C) or a derivative operator (Neumann/Robin B.C). In order to solve the PDE using the PINN method, the residual of the governing PDE is minimized, which is defined by

\begin{equation}
r_{\theta}(\textbf{x})=  \mathcal{N}_{\textbf{x}}[f_{\theta}(\textbf{x})],
\end{equation}
where \(f_{\theta}\) is the predicted value by the network. The residual value, along with the deviation of the prediction from boundary/initial conditions, is used to construct the loss, which takes the form:
\begin{equation}
\label{eqn:PINN_loss}
L(\theta)= L_{r}(\theta)+ \sum^{M}_{i=1}\lambda_{i} L_{i}(\theta),
\end{equation}
where the index i refers to different components of the loss function, relating to initial conditions, boundary conditions, and measurement/simulation data. \(\lambda_{i}\) refers to the weight coefficient of each loss term. The individual loss terms are constituted as follows:

\begin{equation}
L_{r}= \frac{1}{N_{r}}\sum_{i}^{N_{r}} [r(\textbf{x}_{r}^{i})]^{2}, \text{ }
L_{b}= \frac{1}{N_{b}}\sum_{i}^{N_{b}} [\Phi(\hat{u}(\textbf{x}_{b}^{i}))-g_{b}^{i}]^{2}, \text{ }
L_{d}= \frac{1}{N_{d}}\sum_{i}^{N_{d}} [u(\textbf{x}_{d}^{i})-\hat{u}(x_{d}^{i},t_{d}^{i})]^{2},
\end{equation}

where the subscripts r, b, and d refer to collocation, boundary, initial, and data points, respectively. The loss term \(L(\theta)\) can then be minimized to have the network learn the solution to the PDE described by \ref{equation:PDE_main},\ref{equation:PDE_BC}. A popular method is to use gradient-based optimizers like Adam \cite{adam} and L-BFGS to optimize the network weights.

\subsection{Current Challenges with PINNs}
\label{sec:challenges_pinns}
Although the PINN method shows great promise, it still has a number of unresolved issues. The biggest challenges with PINNs currently lie in the scalability of the algorithms to large 3D problems as well as problems with complex nonlinearities, and unsteady problems. Some of the issues described henceforth are tackled by methods described in Section \ref{sec:impfeatures}. 

\subsubsection{Weak imposition of Boundary Conditions}
\label{subsec:weak_imp_bdry}
The solution of a PDE problem must obey all initial and boundary conditions imposed on it while minimizing the residual of the governing equation. However, for neural network based solvers it is difficult to impose boundary and initial conditions in an exact manner. This is because the standard way to impose B.C in PINNs is to create a linear combination of loss functions (as described mathematically in the previous section). Each loss either describes the deviation of the network output from a specific boundary condition, or the magnitude of the residual of the governing equations. Therefore, boundary conditions are only satisfied in a weak manner. There has been research demonstrating the utility of exact imposition of boundary conditions \cite{Sun_2020, Sukumar_2022,hard_constraint_3} or creative multi-network approaches \cite{multi_network_approach}, such implementations are mostly problem-specific and do not generalize well. 

 Weak imposition of boundary conditions also creates another issue, one that is fairly common in multi-task learning and multi-objective optimization: choosing the values of loss term coefficients that make up the linear combination. Choosing these weights is a nontrivial exercise that would require calibration via hyper-parameter search, which is not feasible. \citet{lr_annealing} introduced a heuristic dynamic weighting algorithm to update and select these weights automatically and continuously during the training, to enable convergence to the correct answer. Additionally, there have been several other algorithms proposed to choose the correct scheme for weighting the losses \cite{NTK_1, NTK_2, ReLoBralo}. This continues to be an active area of research in the PINNs community. Finally, methods have been proposed to impose the boundary conditions in a strong manner by manipulating the output formulations \cite{Sun_2020} or by utilizing operator networks \cite{Saad2023}. 

\subsubsection{Difficult Optimization Problem}
A second problem is the nature of the loss landscape itself, in which a reasonable local minimum is required to be found. As seen in \citet{mahoney_PINN}, \citet{gopakumar2022loss},\citet{subramanian2022adaptive} and \citet{Basir_2022}, as well as the author's own experiments, different non-dimensional quantities (e.g. Reynolds number) in the governing equations, the number of dimensions of the problem, the point cloud/discretization, the boundary conditions and the complexity of the solution to be predicted can adversely affect the loss landscape of the neural network training. This makes the optimization challenging and can fail to find an adequate local minimum via a gradient descent-based algorithm. Recently, methods borrowing concepts from optimization theory have shown alternate formulations (e.g. augmented lagrangian method for the loss functions) can aid the convergence properties of the training problem \cite{Basir_2022, son2023enhanced}. There have also been efforts towards imposing physical constraints in an integral form \cite{Hansen2023}.

\subsubsection{Cost of training}
\label{subsubsec:cost_of_training}
Constructing the PDE loss functions involves several backward passes through the network, which is a costly operation. PINNs on average take longer to train than their data-driven counterparts for exactly this reason; the computation graph of a PINN training is much more complex. Moreover, for the Navier-Stokes equations, it has been seen that although the stream function formulation provides better results (due to exact enforcement of continuity), it is costlier in terms of training time. As seen in NVIDIA's experiments \cite{SimNet}, it can take several million iterations for the more complex problems to be solved via PINNs. To reduce the cost of training approaches such as automatic differentiation for finite difference formulations \cite{he2020unsupervised}, or using first-order formulations \cite{gladstone2022fo}, have been proposed. However, these solutions tend to be mostly problem-specific and do not necessarily generalize well to increased problem complexity and grid definitions. Meta-learning algorithms \cite{finn2017model} have also recently gained significance as an effective way to reduce the cost of training neural networks on new tasks, and some of this work has been extended to PINNs \cite{penwarden2023metalearning} as well.

\section{Important Features for Creating PINN Models}
\label{sec:impfeatures}

In this section, the important techniques used to create PINN-based models cost-effectively are outlined. The PINN models in subsequent sections are created by combining these features that have been found to have an effect on the accuracy of the model and the speed of training.

\subsection{Hybrid Data-Physics Training}
\label{subsec:physics_is_better}

\begin{figure}[h!]
    \centering
    \includegraphics[angle=0,width=10 cm]{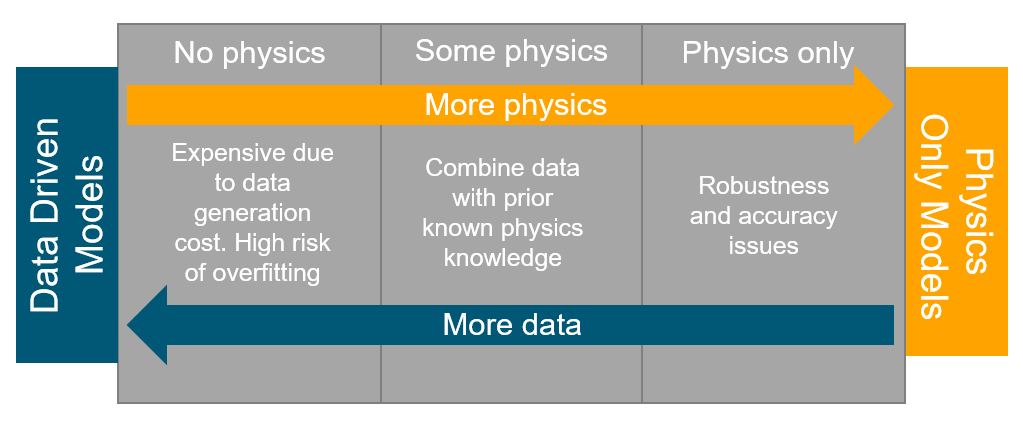}
    \caption{\centering The spectrum of data-driven versus physics-informed models. Incorporating governing physics information into the models during creation serves as an effective form of regularization and often helps reduce the amount of data required to achieve the same accuracy levels.}
    \label{fig:data_physics_tradeoff}
    \centering
    
\end{figure}

Compared with the original PINNs method proposed by \citet{PINN_orig}, a plethora of research has been undertaken to improve and expand on the method \cite{pinn_review_karniadakis,cuomo2022scientific}. From these developments, the PINNs method has been applied to solve PDE-based problems of increasing complexity and dimensionality. However, the PINNs method is currently not suited for solving engineering problems often encountered in industry in a data-free manner. The optimization issues and cost of model training outlined above make the method, presently, unsuitable for use as a forward solver.
To get the best of both worlds, the PINNs method can be augmented with data. Figure \ref{fig:data_physics_tradeoff} depicts the tradeoff between using only data or only physics, and that the sweet spot lies in using both. In addition to the discussed benefit of hybrid data-physics training reducing the cost of generating data, there have been several examples showing that the inclusion of sparse solution data in the training loss function significantly improves the convergence capabilities of the PINNs method \cite{SHARMA2023,pinn_review_karniadakis,gopakumar2022loss}.

In this paper, we take inspiration from this and use very sparse solution data to solve 3D flow-thermal problems and inform our surrogate models with physics while creating them.

\subsection{Modified Learning Rate Annealing}

As described in Section \ref{subsec:weak_imp_bdry}, the learning rate annealing algorithm has proved to be very effective in mitigating the stiffness of the PINN training problem. However, utilizing this method over a broader spectrum of problems highlighted an issue with stability. The following outlines this issue:

As shown in Equation \ref{eqn:PINN_loss}  the PINN loss function being optimized takes the form:

\begin{equation}
L(\theta)= L_{r}(\theta)+ \sum_{i=1}^{M} \lambda_{i}L_{i}(\theta)
\end{equation}

At any training step, the update to the loss coefficient is calculated \cite{lr_annealing} as

\[ \hat\lambda_{i}=\frac{max_{\theta}{|\nabla_{\theta}L_{r}(\theta)|}}{\overline{{|\nabla_{\theta}L_{i}(\theta)|}}}, \space i=1,....,M\]

It can be seen that if the loss \(L_{i}\) decreases much faster than \( L_{r}\) during the training, the value of \(\hat \lambda_{i}\) increases. This then leads to a larger coefficient for that loss term and an associated faster decay of the loss.

This instability has the unintended consequence of the optimizer getting stuck in minima where it minimizes the loss \(L_{i}\) very well but is unable to optimize for the loss of the other constraints. The proposed updated algorithm to mitigate this issue is shown in Algorithm \ref{algo_modded_lr}. The values of thresholds are hyper-parameters, but if the inputs and outputs of the network have been normalized (using standard score normalization, for example), then selecting values between \(10^{-3}\) and \(10^{-5}\) works well in practice.

\begin{algorithm}
    \caption{Modified Learning Rate Annealing}\label{algo_modded_lr}
    \begin{algorithmic}
        
        \FOR{update step = $1$ to $N$}

                \IF{$L_{i}(\theta) \leq (threshold)_{i} $}
                    \STATE $$\hat \lambda_{i}=0$$
                \ELSE
                    \STATE Compute $\hat \lambda_{i}$ by 
                    \STATE $$\hat\lambda_{i}=\frac{max_{\theta}{|\nabla_{\theta}L_{r}(\theta)|}}{\overline{{|\nabla_{\theta}L_{i}(\theta)|}}}, \space i=1,....,M$$
                \ENDIF
                \STATE Update weights $\lambda_{i}$ as 
                \STATE $$\lambda_{i}=(1-\alpha)\lambda_{i}+\alpha \hat \lambda_{i}$$
                \STATE Update network parameters via gradient descent:

                \STATE $$\theta_{n+1}=\theta_{n}- \eta\nabla_{\theta}L_{r}(\theta)-\eta\sum_{i=1}^{M}\lambda_{i}\nabla_{\theta}L_{i}(\theta)$$

        \ENDFOR
        \STATE We set the hyper-parameter \(\alpha=0.1\) and \(\eta=10^{-3}\). Threshold values are chosen somewhere between \(10^{-3}\) and \(10^{-5}\).
    \end{algorithmic}
\end{algorithm}

\begin{figure}[h!]
\centering
\begin{subfigure}{0.45\textwidth}
    \includegraphics[width=\textwidth]{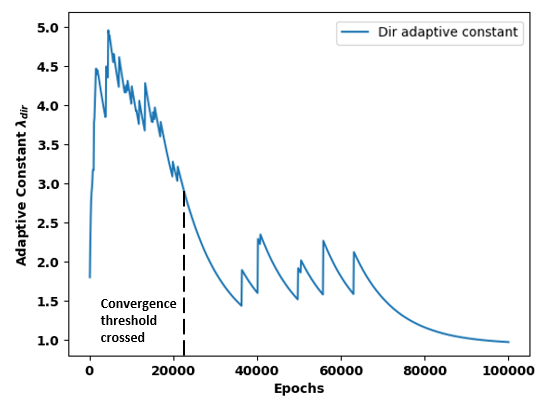}
    \caption{}
    \label{fig:modded_lr_dir_const}
\end{subfigure}
\hfill
\begin{subfigure}{0.45\textwidth}
    \includegraphics[width=\textwidth]{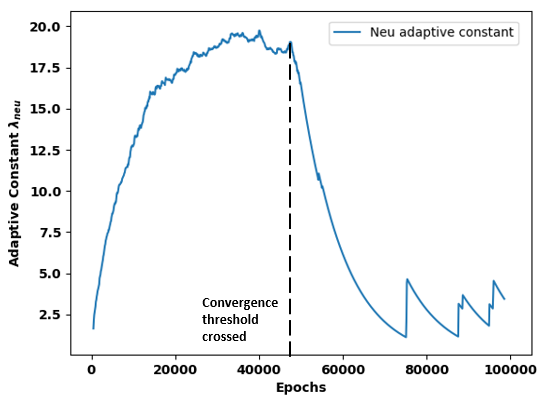}
    \caption{}
    \label{fig:modded_lr_neu_const}
\end{subfigure}
\hfill
\begin{subfigure}{0.45\textwidth}
    \includegraphics[width=\textwidth]{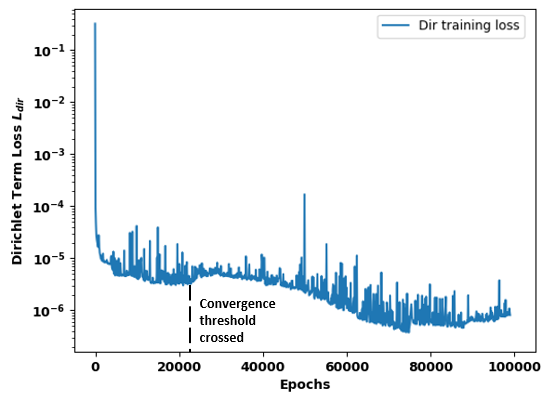}
    \caption{}
    \label{fig:modded_loss_dir_loss}
\end{subfigure}
\hfill
\begin{subfigure}{0.45\textwidth}
    \includegraphics[width=\textwidth]{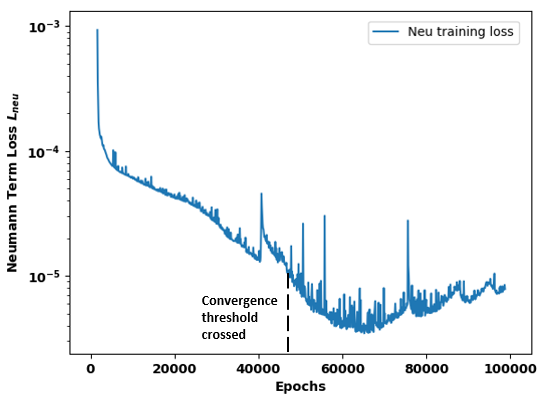}
    \caption{}
    \label{fig:modded_loss_neu_loss}
\end{subfigure}
\hfill
\begin{subfigure}{0.45\textwidth}
    \includegraphics[width=\textwidth]{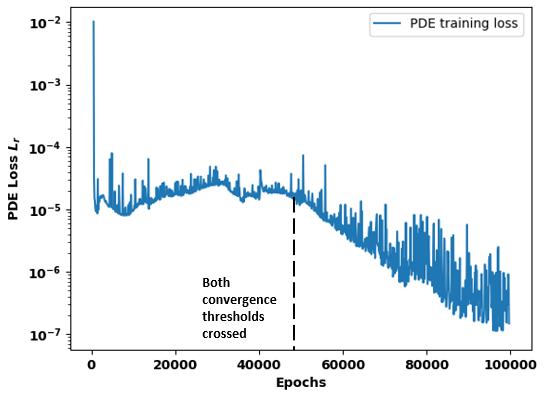}
    \caption{}
    \label{fig:modded_loss_pde_loss}
\end{subfigure}
\hfill
\caption{\centering Adaptive coefficients and loss terms from Equation \ref{eqn:lr_annealing_loss} during training.   
 (\subref{fig:modded_lr_dir_const}) Evolution of the Dirichlet loss adaptive constant during training.(\subref{fig:modded_lr_neu_const}) 
Evolution of the Neumann loss adaptive constant during training.
(\subref{fig:modded_loss_dir_loss})
Dirichlet B.C loss term \(L_{dir}(\theta)\)
(\subref{fig:modded_loss_neu_loss})
Neumann B.C loss term \(L_{neu}(\theta)\)
(\subref{fig:modded_loss_pde_loss})
The PDE loss during training. Once the values of both the adaptive constants start dropping, the PDE loss improves much more rapidly. }
\label{fig:plots_of_coeff}
\end{figure}

For a problem with the loss function 
\begin{equation}
    \label{eqn:lr_annealing_loss}
    L(\theta)=L_{r}(\theta)+\lambda_{neu}L_{neu}(\theta) +\lambda_{dir}L_{dir}(\theta)
\end{equation}

where \(L_{r}(\theta)\), \(L_{neu}(\theta)\) and \(L_{dir}(\theta)\) correspond to the PDE, Neumann and Dirichlet loss respectively, Figure \ref{fig:plots_of_coeff} shows the training curves for the individual losses, and the value of the adaptive coefficients when they are calculated using Algorithm \ref{algo_modded_lr}. It can be seen that when the boundary loss terms in Figures \ref{fig:modded_loss_dir_loss} and \ref{fig:modded_loss_neu_loss} go below their thresholds (set to \(10^{-5}\)), the associated coefficients shown in Figures \ref{fig:modded_lr_dir_const} and \ref{fig:modded_lr_neu_const} start decaying. Following this, the PDE loss starts improving much faster. If the term \(L_{i}(\theta)\) goes above its threshold, it leads to a spike in the adaptive constant \(\lambda_{i}\) which brings it down again.

\subsection{Fourier Feature Embeddings}
\label{sec:fourer_feats}
As described in \citet{tancik2020fourier}, Artificial Neural Networks suffer from a spectral bias problem. To overcome this, they introduced a Fourier feature embedding that allows models to capture high-frequency components of the solution effectively. This has the effect of markedly improving the ability of the networks to capture sharp gradients in the solutions, which requires the network to be able to learn high-frequency components of the solution quickly.

Following the implemenation in \citet{tancik2020fourier}, for an input vector

\[ \mathbf{v}= \left[ {\begin{array}{cc}
   x  \\
   y  \\
   z \\
  \end{array} } \right]\]

instead of using \textbf{v} as the input we compute the Fourier feature mapping:

\begin{equation}
 \gamma(\mathbf{v})=[\cos(2\pi \textbf{b}^{T}_{1}\textbf{v}),\sin(2\pi \textbf{b}^{T}_{1}\textbf{v}),.....,\cos(2\pi \textbf{b}^{T}_{m}\textbf{v}),\sin(2\pi \textbf{b}^{T}_{m}\textbf{v})]
\end{equation}
where m is a hyper parameter and the frequencies \(\textbf{b}^{T}_{j}\) are selected randomly from an isotropic distribution. Then $\gamma(\mathbf{v})$ is passed into the network.

The Fourier feature embedding was shown to be highly effective in training PINNs models by \citet{wang2021eigenvector}, and several results were shown for 1D and 2D problems. We extend this implementation to solve 3D flow problems via PINNs and use it to create our hybrid data-PINN surrogate for flow thermal problems.

In addition, there have been other proposed solutions for the spectral bias problem for applications to PDE problems, such as the Siren activation \cite{siren}, Fourier Neural Operators \cite{FNO}, and weighting schemes derived from the theory of Neural Tangent Kernels (NTK) \cite{NTK_1}.

\section{Experiments and Results}
\label{sec:exp_and_results}
In this section, some example problems are solved using PINNs. Sections \ref{sec:stenosis_problem} and \ref{sec:pcb_solve} solve the 3D incompressible Navier-Stokes equations through a data-assisted approach, where very sparse solution data is provided in the domain.

Section \ref{sec:heat_sink_design} uses a hybrid data-PINN approach to generate a surrogate model for a given design space of a heat sink with a chip underneath it, undergoing cooling via forced convection. Then, given certain constraints on the running metrics of the chip-sink setup (like max temperature in the chip), the optimal set of parameters in the Design of Experiments (DoE) space that satisfy the constraints while maximizing an objective are obtained via rapid design optimization using the created surrogate.

Details on hyper-parameters used in the model training for each experiment that follows can be found in Appendix Section \ref{sec:pinn_setup_training}.

\subsection{Forward Solve of 3D Stenosis Problem}\label{sec:stenosis_problem}

\begin{figure}[h!]
    \centering
    \includegraphics[angle=0,width=10 cm]{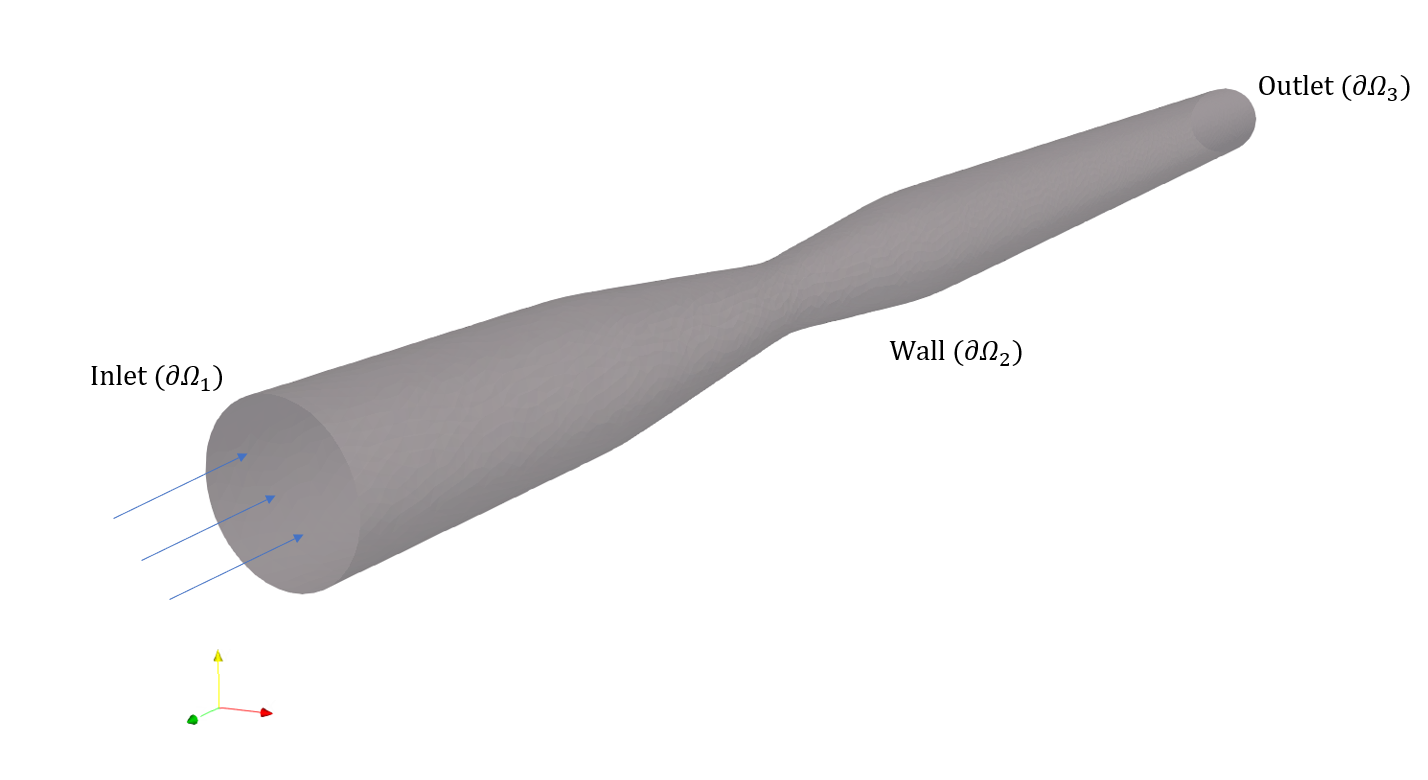}
    
    \caption{Visual description of stenosis problem}
    \label{fig:geo_stenosis}
    \centering
\end{figure}

 Flow through an idealized 3D stenosis geometry at a physiologically relevant Reynolds number is demonstrated, see Figure \ref{fig:geo_stenosis} for details about the geometry. To the author’s best knowledge, flow through a stenosis has been solved using PINNs only at a low Reynolds number of approximately 6 (based on inlet diameter) \cite{Sun_2020}. Flow through irregular geometries has been solved at a higher Re (500), but in 2D \cite{geo_pinn}.  
In this paper, the stenosis problem is solved at Re 150, and in 3 dimensions. 

\begin{figure}[h!]
\centering

    \includegraphics[width=1.0\textwidth]{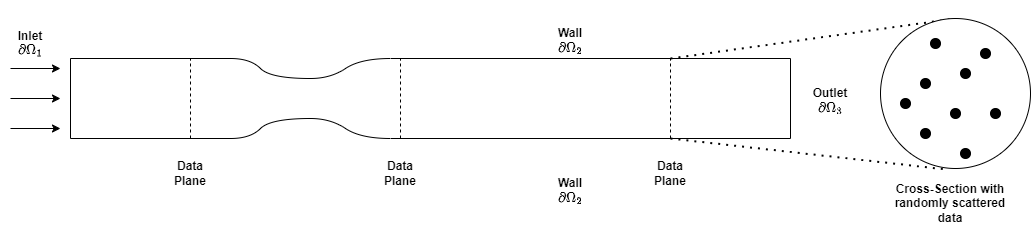}

\hfill
        
\caption{\centering Stenosis diagram (not to scale) showing planes where solution data is provided randomly. }
\label{fig:data_pts_stenosis}
\end{figure}

As discussed in Section \ref{sec:challenges_pinns}, at higher Reynolds numbers the standard PINN implementation struggles to achieve a good local minimum. This was confirmed using a standard PINN implementation. To alleviate this issue a data-assisted approach where sporadic solution data can be added throughout the domain of interest (depicted on a slice in Figure \ref{fig:data_pts_stenosis}). The data was given in the form of concentric rings at the radii depicted on the cut plane.

 \subsubsection{Problem Setup}
 
  The flow problem through the stenosis is solved by solving the steady-state incompressible Navier-Stokes equations:

\begin{equation}
\nabla \cdot \textbf{u}=0,
\end{equation}
\begin{equation}
(\textbf{u} \cdot \nabla )\textbf{u}=-\frac{1}{\rho}\nabla\textbf{p}+\nu\nabla\cdot(\nabla\textbf{u}),
\end{equation}
subject to

\[ \textbf{u}(x_{b1})=g(x_{b1})\text{, } x_{b1} \in \partial \Omega_{1}, \]

\[\textbf{u}(x_{b2})=0\text{, } x_{b2} \in \partial \Omega_{2}, \]

\[\nabla u_{i}(x_{b3})\cdot \textbf{n}=0\text{, } x_{b3} \in \partial \Omega_{3} , i=1,2,3 \]

\[p(x_{b3})=0 \text{, } x_{b3} \in \partial \Omega_{3}\]

where \(g(x_{b3})\) represents a profiled input to the stenosis. \(\rho\) and \(\nu\) are the density and kinematic viscosity of the fluid(air) respectively, and \textbf{u} and p are the velocity vector and pressure respectively. \\

In the present problem, a parabolic profiled input is provided with a peak value inflow of 0.15 m/s. The ratio of areas of the throat to the inlet is 0.36.

 The output of the network is approximated as \( G_{\theta} \), which is a 4-component output:

\[
  G_{\theta}=
  \left[ {\begin{array}{cc}
   u \\
   v \\
   w \\
   p \\
  \end{array} } \right]
\]

\subsubsection{Results}

\begin{figure}[h!]
\centering
\begin{subfigure}{0.7\textwidth}
    \includegraphics[width=\textwidth]{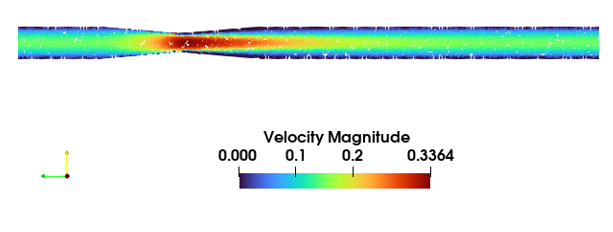}
    \caption{}
    \label{fig: acuSolve_stenosis_contours}
\end{subfigure}
\hfill
\begin{subfigure}{0.7\textwidth}
    \includegraphics[width=\textwidth]{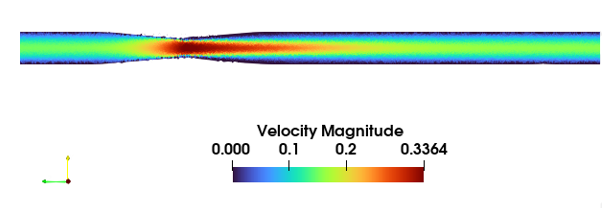}
    \caption{ }
    \label{fig:PINN_stenosis_contours}
\end{subfigure}
\hfill
        
\caption{\centering Solution Comparison. (\subref{fig: acuSolve_stenosis_contours}) Altair AcuSolve\textsuperscript{\textregistered} Solution to stenosis problem (\subref{fig:PINN_stenosis_contours}) PINN forward solve to stenosis problem. }
\label{fig:stenosis_diff}
\end{figure}

\begin{figure}[h!]
\centering
\begin{subfigure}{0.45\textwidth}
    \includegraphics[width=\textwidth]{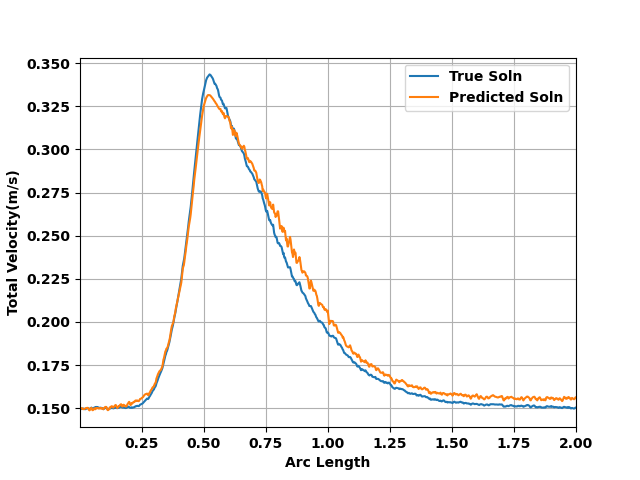}
    \caption{}
    \label{fig:vel_comp_stenosis}
\end{subfigure}
\hfill
\begin{subfigure}{0.45\textwidth}
    \includegraphics[width=\textwidth]{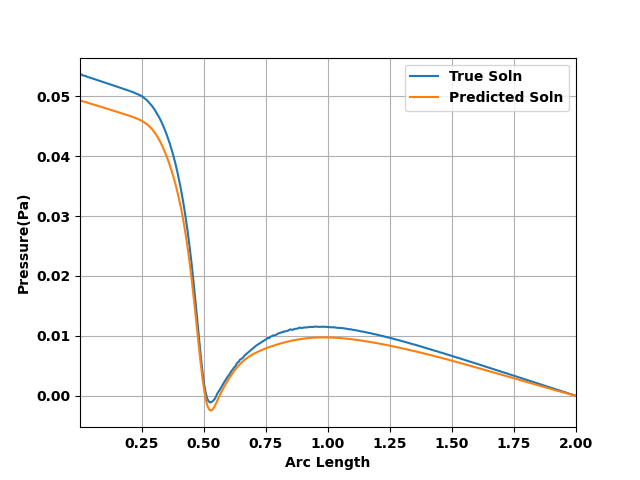}
    \caption{}
    \label{fig:pres_comp_stenosis}
\end{subfigure}
\hfill
        
\caption{\centering Centerline solution comparisons: PINN versus Altair AcuSolve\textsuperscript{\textregistered} (\subref{fig:vel_comp_stenosis})  Total Velocity Comparison (\subref{fig:pres_comp_stenosis}) Pressure Comparison}
\label{fig:line_plots_stenosis}
\end{figure}

Figure \ref{fig:stenosis_diff} compares the velocity magnitude returned by the trained PINN model and Altair AcuSolve\textsuperscript{\textregistered} through a 2D slice of the stenosis. As can be seen, the essential features of the flow are captured. Figure \ref{fig:vel_comp_stenosis} and \ref{fig:pres_comp_stenosis} compare the velocity and pressure profile through the center of the stenosis. The differences between the line plots are attributed to differences in mesh density between the two cases. The CFD mesh was an unstructured mesh of around 63,000 nodes with a boundary layer, while the point cloud used with the PINN was around 87,000 nodes randomly distributed points except near the boundary where the sampling was finer.

Another approach that was investigated to solve the 3D stenosis problem was that of using "continuity planes" as defined by \citet{SimNet} in their experiments solving 3D flow problems using PINNs. In this approach, the authors added constraints on the mass flow through a plane and added these constraints to the loss function. While this approach was found to aid the convergence of the PINN model to the correct solution, there were several issues found to exist with this method:

\begin{enumerate}
    \item It is difficult to generate continuity planes for complex geometries such as those shown in Sections \ref{sec:pcb_solve} and \ref{sec:heat_sink_design}.

    \item The quality of the solution from the PINN depends heavily on the integration scheme used to calculate the mass flow rate, and the fineness of the points on the continuity plane.
\end{enumerate}

Hence, in the next section, random and sparsely distributed data was used in the domain to aid convergence.

\subsection{Flow over a Printed Circuit Board (PCB)}
\label{sec:pcb_solve}

\subsubsection{Problem Setup}

\begin{figure}[h!]
\centering

    \includegraphics[width=0.75\textwidth]{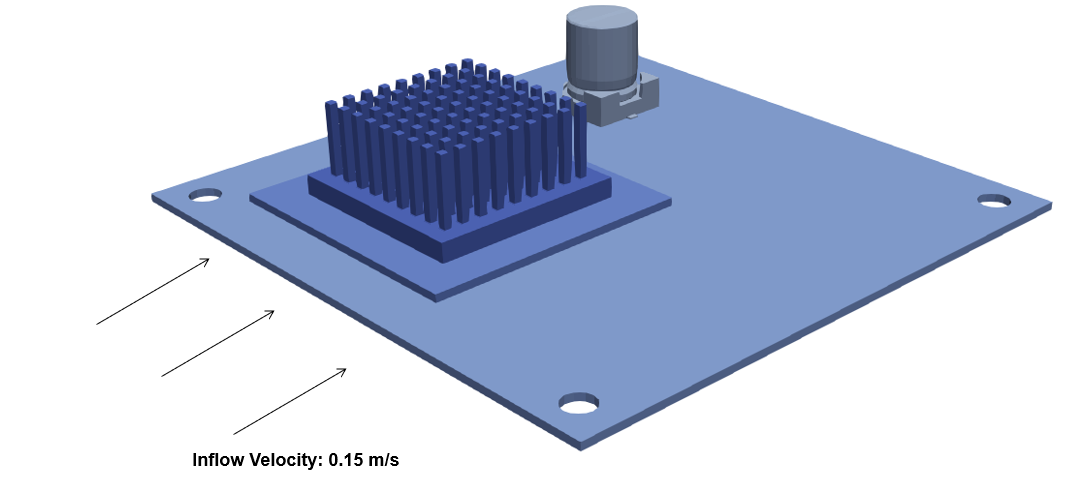}
\hfill
        
\caption{Geometry of a PCB with a chip, sink, and capacitor assembly.}
\label{fig:pcb_flow_desc}
\end{figure}

Flow over a PCB consisting of a heat sink, chip, and capacitor is solved at a Reynolds number of approximately 1500, based on the length of the PCB board and air as the fluid. The geometry and flow orientation are shown in Figure \ref{fig:pcb_flow_desc}. This represents a forced convection problem common in electronics design and is a challenging problem for PINNs because it is in 3D, with a complex geometry and large gradients involved.

Let \(D\) represent the set of all nodes in the domain. To train the PINN model, the CFD solution was first computed. Next, 1\% of the nodes in the solution domain were randomly selected (call this set \(D_{1} \subset D)\)). This is a selection of roughly 2,300 node points (from a mesh of roughly 230,000 nodes). The experiment was then divided into three parts:

\begin{enumerate}
    \item \textbf{Case A}: A network was trained on the CFD solution at all points in \(D_{1}\) (i.e \(\forall \textbf{x} \in D_{1} \)) following which the physics of the problem was \textbf{enforced at every node location} in \(D\) (i.e \(\forall \textbf{x} \in D \)), by including the physics-based loss in the training, and then the network was asked to predict the solution in the entire domain \(D\).
    \item \textbf{Case B}: A network was trained on the CFD solution at the points contained in \(D_{1}\) (i.e, \(\forall \textbf{x} \in D_{1} \)) \textbf{without any physics enforcement} and then asked to predict the solution in the entire domain (i.e \(\forall \textbf{x} \in D \)).
    \item \textbf{Case C}: Finally, the same experiment as Case A was repeated but with a new set \(D_{2}\) consisting of only 0.2\% of the nodes in \(D\), which were again randomly selected.
\end{enumerate}

The governing equations for this problem are the Reynolds Averaged Navier-Stokes Equations:
\begin{equation}
\nabla \cdot \textbf{u}=0,
\end{equation}

\begin{equation}
(\textbf{u} \cdot \nabla )\textbf{u}=-\frac{1}{\rho}\nabla\textbf{p}+(\nu+\nu_{t})\nabla\cdot(\nabla\textbf{u}),
\end{equation} 
\(\rho\), \(\nu\), and \(\nu_{t}\) represent the density, kinematic viscosity and eddy viscosity of the system. The inflow is set to a constant velocity of 0.15 m/s and the outflow is set to the stress-free condition. It should be noted that in the current study, eddy viscosity is obtained directly from the CFD solver using the Spalart-Allmaras turbulence model.  Turbulence modeling in PINNs is a field of active research with a few articles investigating it  \cite{SimNet,RANS_PINN,PINN_turb_modeling}, 
and it is a future work to effectively incorporate turbulence models into PINN-based models.

\subsubsection{Results}

\begin{figure}[h!]
\centering

\begin{subfigure}{0.4\textwidth}
    \includegraphics[width=\textwidth]{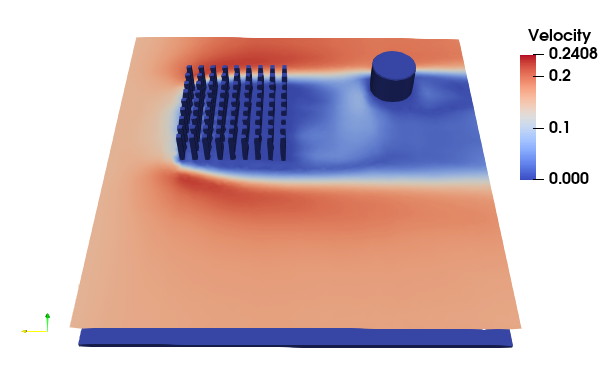}
    \caption{}
    \label{fig:1_perc_data_phys}
\end{subfigure}
\hfill
\begin{subfigure}{0.4\textwidth}
    \includegraphics[width=\textwidth]{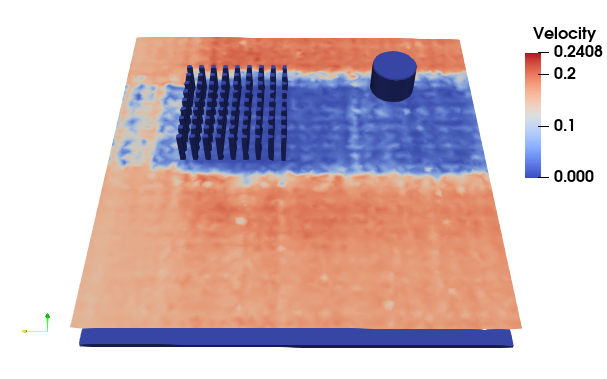}
    \caption{}
    \label{fig:1_perc_data}
\end{subfigure}
\hfill
\begin{subfigure}{0.4\textwidth}
    \includegraphics[width=\textwidth]{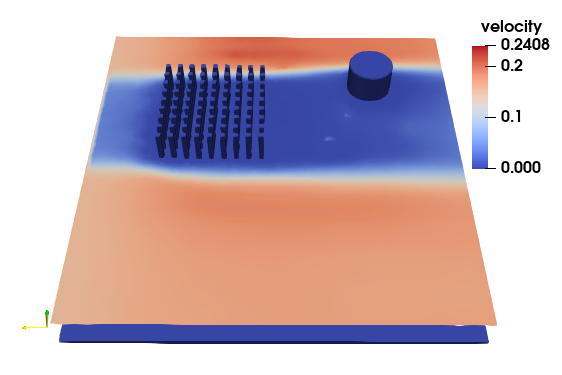}
    \caption{}
    \label{fig:p2_perc_data}
\end{subfigure}
\hfill
\begin{subfigure}{0.4\textwidth}
    \includegraphics[width=\textwidth]{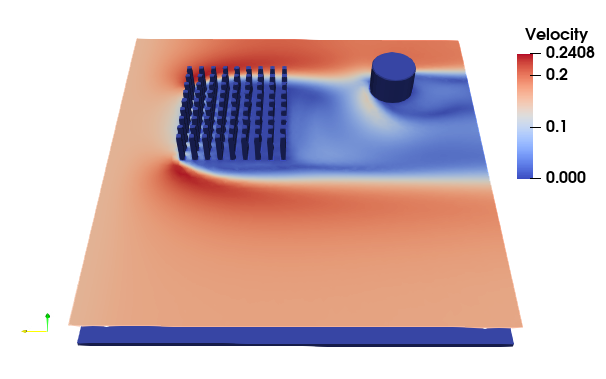}
    \caption{}
    \label{fig:true_sol_1_perc_data}
\end{subfigure}
\hfill
        
\caption{\centering Neural Network (NN) prediction with and without physics, for very coarse data supplied on a plane through the domain. (\subref{fig:1_perc_data_phys}) \textbf{Case A:} Trained on 1\% data and physics
(\subref{fig:1_perc_data}) \textbf{Case B:} Trained on 1\% solution data only
(\subref{fig:p2_perc_data}) \textbf{Case C:} Trained on 0.2\% data and physics
(\subref{fig:true_sol_1_perc_data}) True Solution from CFD solver
}
\label{fig:data_physics_differences}
\end{figure}

Figure \ref{fig:data_physics_differences} shows the ANN predictions for the different cases. It is evident that by using sparse data, the network is able to better converge toward the CFD solution (shown in Figure \ref{fig:true_sol_1_perc_data}) using the physics-based regularizer. However, as evident in Figure \ref{fig:p2_perc_data}, the network failed to converge to a physical solution when the amount of data provided was insufficient, highlighting the importance of a certain amount and fineness of the required data. Table \ref{table:pcb_errors} shows the Mean Squared Errors (MSE) for each experiment, for the velocity and the pressure, taking the CFD solution as the ground truth. The MSE is calculated as

\begin{equation}
\label{eqn:mse_formula}
\text{MSE}=\sqrt{\frac{\sum_{i=1}^{N_{nodes}}(x_{i,pred}-x_{i,truth})^{2}}{N_{nodes}}} 
\end{equation}

\begin{figure}[h!]
\centering
\begin{subfigure}{0.45\textwidth}
    \includegraphics[width=\textwidth]{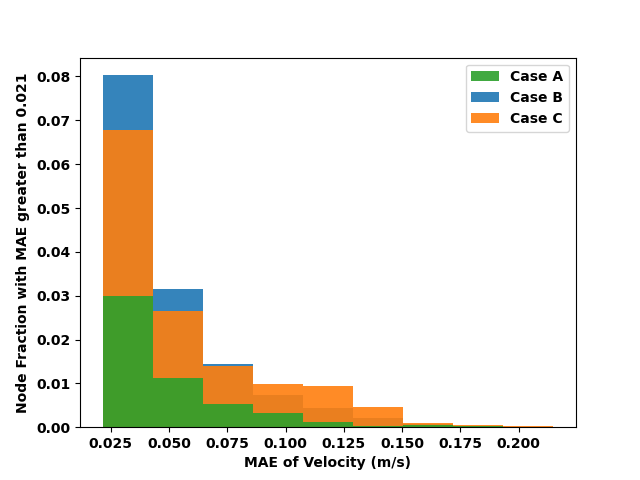}
    \caption{}
    \label{fig:pcb_vel_node_fracs}
\end{subfigure}
\hfill
\begin{subfigure}{0.45\textwidth}
    \includegraphics[width=\textwidth]{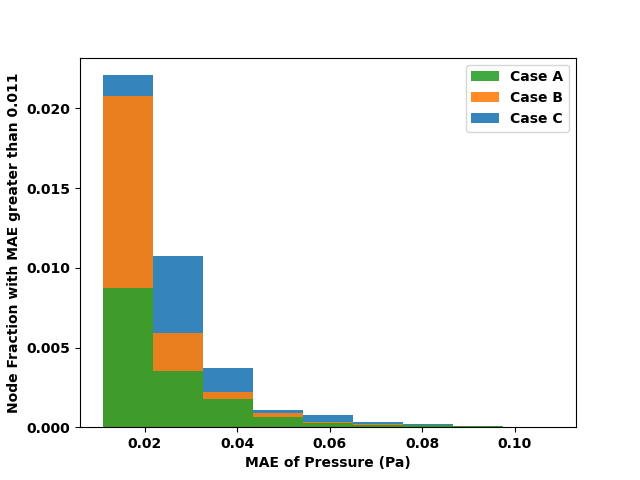}
    \caption{}
    \label{fig:pcb_pres_node_fracs}
\end{subfigure}
\hfill
        
\caption{\centering Node fractions of points above a certain MAE value, for each case. (\subref{fig:pcb_vel_node_fracs})  MAE of Velocity (\subref{fig:pcb_pres_node_fracs}) MAE of Pressure}
\label{fig:mae_stats_pcb}
\end{figure}

Figure \ref{fig:mae_stats_pcb} shows the fraction of node points for each case that are above a certain Mean Absolute Error (MAE) value. The lower the fraction, the better the solution. We note from Figure \ref{fig:mae_stats_pcb} that even for Case A, there are outliers to the solution where the MAE is relatively high, indicating poor convergence to the solution at those nodes. The convergence of PINNs toward the correct solution for highly nonlinear systems is an open and challenging problem, especially in 3 dimensions. Nonetheless, these results open exciting possibilities about using physics-based regularizers in the future and represent a step forward for solving the 3D Navier-Stokes Equations at high Reynolds Numbers using PINNs. Furthermore, data generation costs to create surrogate models using PINNs can be greatly reduced by providing solution data on a coarser grid and solving the physics on a finer grid. 

\begin{table}[h!]
\begin{center}
\begin{tabular}{ |c|c|c|c| } 
 \hline
 Case & Description  & MSE (Velocity)  & MSE (Pressure)  \\ 
 \hline
  Case A & 1\% domain data + physics & \textbf{0.0135} & \textbf{0.0037}  \\ 
  \hline
  Case B & 1\% domain data only & 0.0222 & 0.00472  \\
  \hline
  Case C & 0.2\% domain data + physics & 0.0245 & 0.00545  \\
 \hline
\end{tabular}
\caption{Mean Squared Errors (MSE) for velocity and pressure for cases A,B, and C}
\label{table:pcb_errors}
\end{center}
\end{table}

\subsection{Surrogate Modeling and Design Optimization of a Heat Sink}
\label{sec:heat_sink_design}

\begin{figure}[h!]
  \centering
  \begin{minipage}[b]{0.9\textwidth}
    \includegraphics[width=\textwidth]{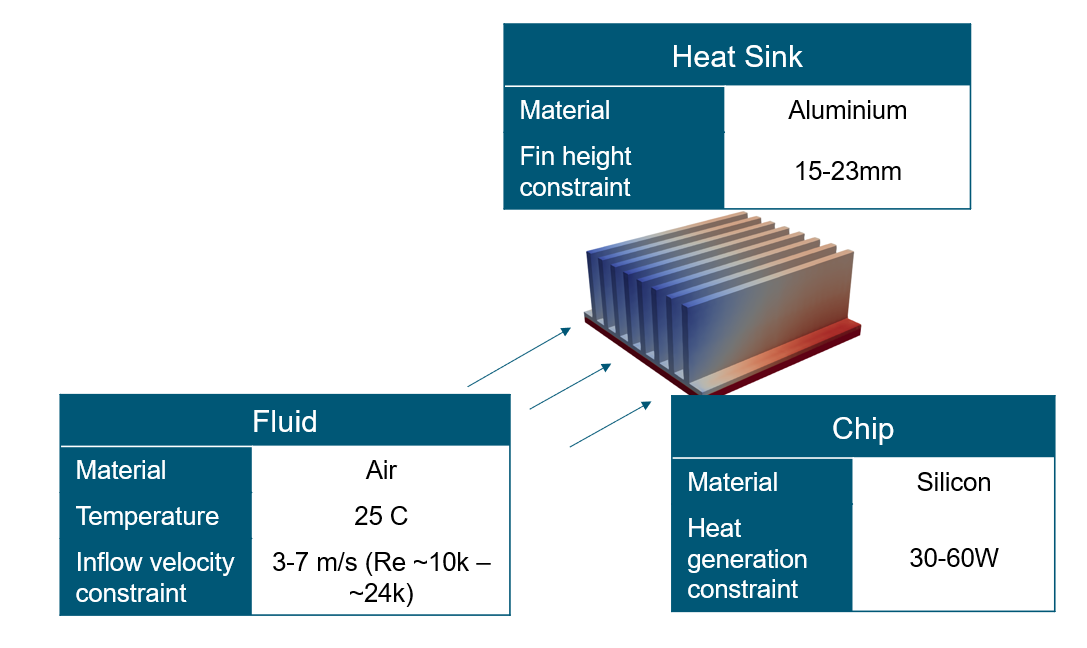}
    \caption{Basic problem geometry and flow depiction}
    \label{fig:heatsink_depic}
  \end{minipage}
  \hfill
  
\end{figure}

In this section, the PINNs surrogate modeling technique is demonstrated for rapid design optimization of a heat sink assembly. The assembly utilizes a chip that generates heat and a fin-type heatsink on top to dissipate heat into the surrounding fluid. The chip-heatsink assembly is cooled by forced convection of air. The geometry and setup are shown in Figure \ref{fig:heatsink_depic}.

The goal is to optimize the heat sink design and the running conditions of the assembly, subject to feasibility constraints placed on chip temperature and channel pressure drop. This represents a common design optimization problem in electronics cooling. More specifically, if \(\dot Q_{src}\) is the total power being generated by the chip, the optimization problem can be framed as follows:

\begin{equation}
\label{eqn:opt_1}
\text{Maximize } \dot Q_{src} \text{ s.t}
\end{equation}

\begin{equation}
\label{eqn:opt_2}
\text{Pressure drop across the heat sink channel \( (\Delta \text{P})\leq\) 11 Pa}
\end{equation}

\begin{equation}
\label{eqn:opt_3}
\text{Maximum temperature anywhere on the chip \(\leq\) 350 K}
\end{equation}

The pressure at the outflow is fixed to 0 Pa, and the pressure drop across the heat sink channel is hence calculated as the average pressure over the inflow of the channel:

\begin{equation}
    \Delta \text{P}= \overline{\text{P}_{\text{inlet}}} 
\end{equation}

The term to be maximized \(\dot Q_{src}\) is also one of the design axes and an input parameter(P3) to the network. 

The design variables that can be altered for this present optimization are: 
\begin{itemize}
    \item Inflow Velocity
    \item Fin height
    \item Source term in the chip (has to be maximized)
\end{itemize}

The upper and lower limits of each of the design variables mentioned above are summarized in Table \ref{table:heat_sink_params}. The inlet velocity is set based on typical values found in literature \cite{lindstedt2012optimization} and corresponds to a Reynolds number range of Re ~10,300 to Re 24,000.

\begin{table}[h!]
\begin{center}
\begin{tabular}{ |c|c|c|c| } 
 \hline
 Parameter No. & Parameter Name  & Lower Value  & Upper Value  \\ 
 \hline
  P1 & Inflow Velocity (\(m/s\)) & 3 & 7  \\ 
  \hline
  P2 & Fin Height (\(mm\)) & 15 & 23  \\
  \hline
  P3 & Source Term (\(W\)) & 30 & 60 \\
 \hline
\end{tabular}
\caption{Design of Experiments space axes ranges for the heat sink design optimization}
\label{table:heat_sink_params}
\end{center}
\end{table}

The governing equations solved for this conjugate heat transfer problem are the same as in Section \ref{sec:pcb_solve} for the flow problem, subject to no-slip boundary conditions on the chip-heatsink assembly with a variable freestream inflow velocity, causing forced convection. As in Section \ref{sec:pcb_solve}, the eddy viscosities are taken from the CFD solutions.

The energy equation in both fluid and solid reads:
\begin{equation}
k\nabla^{2}T + \dot{q}_{src} - \rho s\textbf{u} \cdot \nabla T = 0,
\end{equation}
where T represents the temperature, \(\dot q_{src}\) represents the volumetric source term, and \(k\) and \(s\) are the conductivity and specific heat of the material respectively. At the interface between the fluid and solid domain (fluid-sink, sink-chip, and fluid-chip) the interface condition is applied by minimizing the following loss terms as shown in \cite{cPINN};

\begin{equation}
\label{eqn:interface_1}
L_{flux}=\frac{1}{N_{int}}\sum_{i=1}^{N_{int}}(f_{d_{1}}(\textbf{u}(x_{i}))\cdot \textbf{n}_{d_{1}} + f_{d_{2}}(\textbf{u}(x_{i}))\cdot \textbf{n}_{d_{2}} )^{2}, 
\end{equation}

\begin{equation}
\label{eqn:interface_2}
L_{val}= \frac{1}{N_{int}}\sum_{i=1}^{N_{int}} (\textbf{u}_{d_{j}}(x_{i}) -\overline{\textbf{u}_{d_{j}}(x_{i})} )^{2}, 
\end{equation}

where \(\textbf{n}_{d1}=-\textbf{n}_{d2}\) and j=1,2. The average is taken over j. \(d_{1}\) and \(d_{2}\) refer to the domains on both sides of the interface, and \(N_{int}\) is the number of node points on the interface.

\paragraph{Model Creation and Evaluation} \mbox{} \\

\begin{figure}[h!]
  \centering
  \begin{minipage}[b]{0.7\textwidth}
    \includegraphics[width=\textwidth]{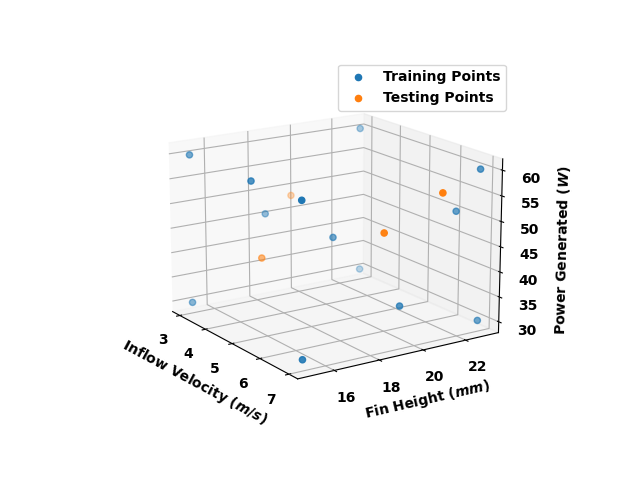}
    \caption{Training and testing points in the 3D DoE space}
    \label{fig:doe_depic}
  \end{minipage}
  \hfill
  
\end{figure}

The sampling of the above Design of Experiments (DoE) space is done via an efficient space-sampling method to optimally fill the DoE space \cite{morris_mitchell}. The sampled DoE space for training is shown in Figure \ref{fig:doe_depic}, along with the location of points at which the surrogate model is tested. The reader is referred to Section \ref{sec:doe_tables} for a complete tabular description of the DoE space. Note that for this example, we use full field data at each DoE point to train the surrogate as opposed to a small fraction of it (like in Section \ref{sec:stenosis_problem} and \ref{sec:pcb_solve}), as the objective is to get a surrogate that is as accurate as possible. Table \ref{table:surrogate_test_pt_errors} shows the MSE for the predictions by the hybrid data-PINN model at the test points, calculated w.r.t the CFD solution at the same mesh resolution. Also shown is the MSE for predictions by a standard data-driven NN without leveraging key features described in Section \ref{sec:impfeatures}, which are used extensively in industry for surrogate modeling applications. The hybrid data-PINN model outperforms the standard data-driven NN for all predictions. Section \ref{sec:imp_above_baseline} shows some more qualitative comparisons between test point results from the PINNs model versus standard data-driven NNs.

\begin{table}
\centering
\begin{tabular}{@{}lccc@{}}
\toprule
     & \textbf{Velocity MSE} & \textbf{Pressure MSE}  & \textbf{Temperature MSE} \\ 
\midrule
 \textbf{Test Point 1}\\
    Hybrid data-PINN Model  & \textbf{0.65}    & \textbf{2.62}    & \textbf{1.81}    \\
    Standard Data-driven NN    & 0.93    & 2.83     & 2.05     \\
    
  \midrule
     \textbf{Test Point 2}\\
    Hybrid data-PINN Model  &  \textbf{0.39}   &  \textbf{1.19}   &  \textbf{2.67}  \\
    Standard Data-driven NN   &   0.58  &  1.42   & 2.97   \\
    
  \midrule
   \textbf{Test Point 3}\\
    Hybrid data-PINN Model   &   \textbf{0.76}  &  \textbf{3.31}   &  \textbf{1.86}  \\
    Standard Data-driven NN     &  1.10   &  3.51   &  2.18  \\
  \midrule
   \textbf{Test Point 4}\\
    Hybrid data-PINN Model   &  \textbf{0.33}   &  \textbf{0.99}   &  \textbf{2.87}  \\
    Standard Data-driven NN     &   0.52  &  1.19   &  3.15  \\  
  \bottomrule                          
\end{tabular}
\caption{\centering MSE for 4 test points shown in Table \ref{table:surrogate_testing_table}. The PINN-based model consistently outperforms the standard data-driven NN on all test points.}
\label{table:surrogate_test_pt_errors}
\end{table}

\paragraph{Solving the Design Optimization Problem} \mbox{} \\

The surrogate model is used to solve the design optimization problem described in Equations \ref{eqn:opt_1}-\ref{eqn:opt_3}. The goal is to show that the surrogate model can accurately predict the solution in the entire DoE space by returning a solution that satisfies all applied constraints while maximizing the objective. The created surrogate models are interfaced with an optimizer that solves a generic constrained optimization problem via an iterative process, described thoroughly in Appendix Section \ref{subsec:PSO}. 

\begin{figure}[h!]
\centering
\begin{subfigure}{0.3\textwidth}
    \includegraphics[width=\textwidth]{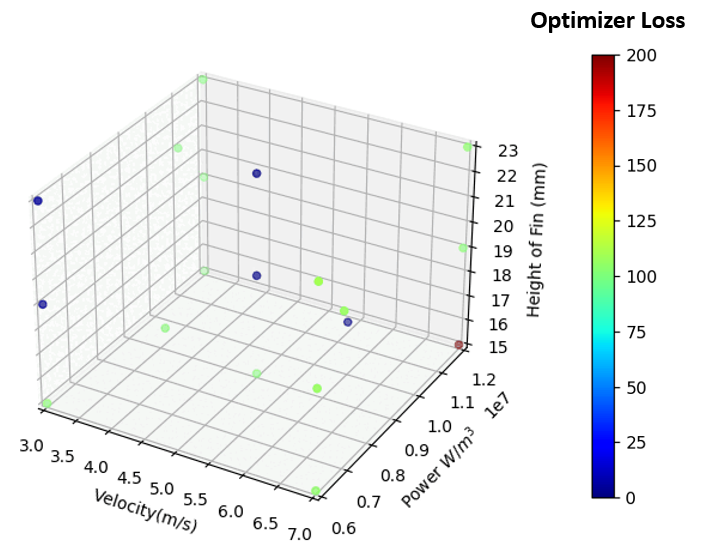}
    \caption{}
    \label{fig:iter0_sink}
\end{subfigure}
\hfill
\begin{subfigure}{0.3\textwidth}
    \includegraphics[width=\textwidth]{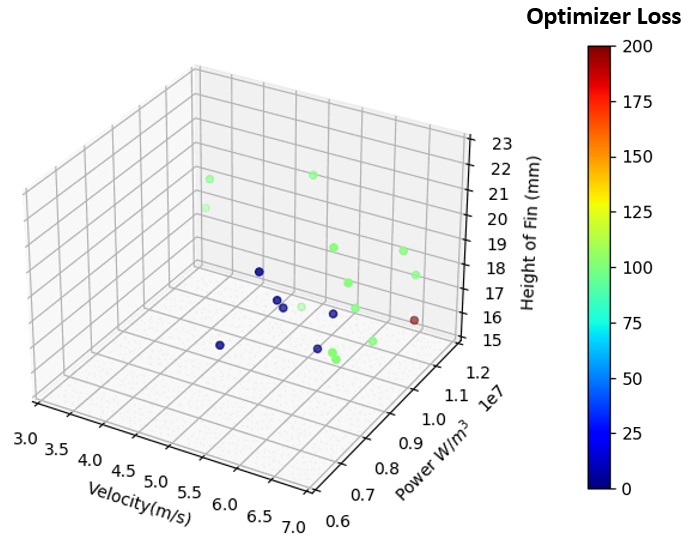}
    \caption{}
    \label{fig:iter5_sink}
\end{subfigure}
\hfill
\begin{subfigure}{0.3\textwidth}
    \includegraphics[width=\textwidth]{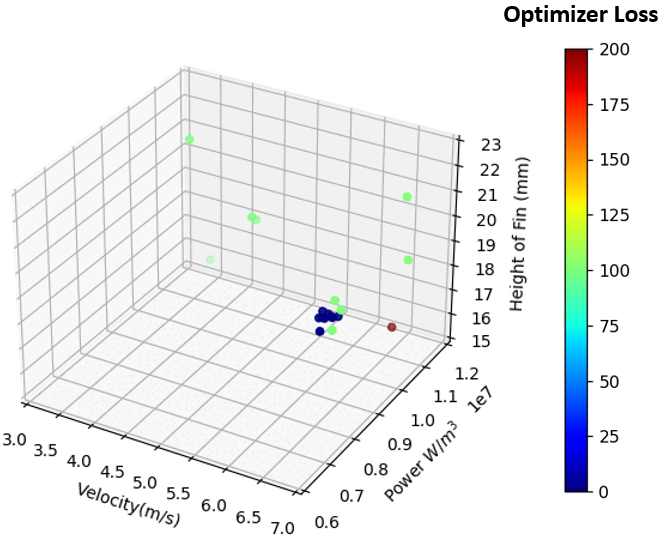}
    \caption{}
    \label{fig:iter10_sink}
\end{subfigure}
        
\caption{\centering Design optimization iterations of the heat sink problem (\subref{fig:iter0_sink}) Iteration 0 (\subref{fig:iter5_sink}) Iteration 5
(\subref{fig:iter10_sink}) Iteration 10}
\label{fig:heat_sink_PSO}
\end{figure}

Each snapshot in Figure \ref{fig:heat_sink_PSO}  represents a design iteration, and each particle represents a point in the DoE space. Each axis of a plot represents a parameter axis.

For the given constraints, the particles converge to a much smaller region of the DoE space. The design point returned by the optimizer in this case is:

\begin{center}
    
\textbf{Inflow Velocity}: 6 m/s \\
\textbf{Chip Power}: 50W\\
\textbf{Fin Height}: 17mm\\

\end{center}

\begin{figure}[h!]
  \centering
  \begin{minipage}[b]{0.9\textwidth}
    \includegraphics[width=\textwidth]{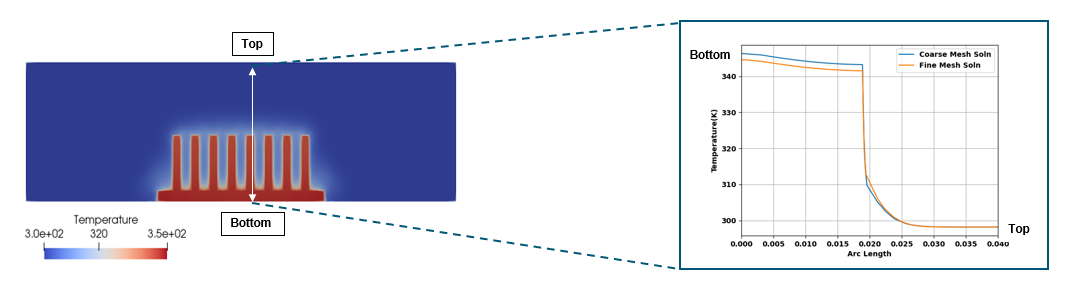}
    \caption{\centering Temperature plot through a slice at the rear of the sink (from bottom to top). The comparison between the high-fidelity solution on the fine mesh and the PINN prediction on a coarser mesh shows good agreement.}
    \label{fig:sink_thermal}
  \end{minipage}
  
  \hfill 
\end{figure}

\begin{figure}[h!]
  \centering
  \begin{minipage}[b]{0.9\textwidth}
    \includegraphics[width=\textwidth]{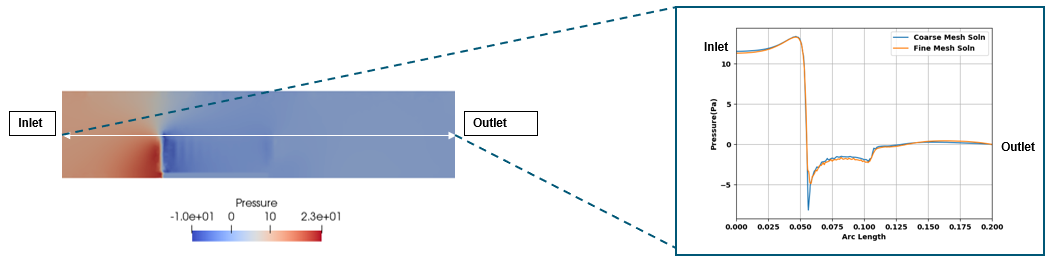}
    \caption{\centering Pressure plot through a slice through the middle of the flow channel (from left to right). The comparison between the high-fidelity solution on the fine mesh and the PINN prediction on a coarser mesh shows good agreement}
    \label{fig:sink_pressure}
  \end{minipage}
  
  \hfill 
\end{figure}

To test that the result satisfies the constraints, the returned design point is solved by the Altair AcuSolve\textsuperscript{\textregistered}, at the same mesh fineness and another mesh with 10x fineness, keeping all essential mesh features such as boundary layers and refinement zones. As shown in Figures \ref{fig:sink_thermal} and \ref{fig:sink_pressure}, not only does the given design point satisfy the design constraints, but the finer mesh solution is very close to the coarser solution, and a little tweaking of the design point using CFD with the higher resolution mesh will yield a highly optimized solution to the designer. This optimization is done several orders of magnitude faster than if using traditional CFD, and the reader is referred to Appendix Section \ref{sec:cost_benefit_analysis_heat_sink} for a quantitative description of the same.

\section{Conclusions and Future Work}
\label{sec:conclusions}
In this paper, Physics Informed Neural Networks were used to solve the 3D Navier-Stokes equations in a data-assisted setting, for complex geometries with realistic physical parameters. It was shown that even for problems being solved at high Reynolds Numbers in 3D, PINNs can be trained to produce a good solution in the presence of very sparse solution data randomly scattered in the solution domain. However, using too little solution data causes the model to converge to an unphysical solution. PINNs were also demonstrated for 3D flow-thermal surrogate modeling and the PINN-based surrogates consistently outperformed standard data-driven NN on test case examples. The PINN surrogates were also interfaced with a design optimization algorithm to solve a constrained optimization problem. This optimization returned a design point that when solved with high-fidelity CFD was consistent with the requirements of the design constraints, highlighting the suitability of the method to produce surrogates for 3D flow-thermal surrogate modeling problems.\\

There are multiple avenues through which the work shown in this paper can be improved. Research has to be done to improve convergence and offer guarantees of PINNs training toward the local minimums that represent physical solutions, in a data-free manner. This will further reduce data requirements for the creation of physically consistent PINN models which can greatly improve their surrogate modeling capabilities, by reducing the cost of training and improving predictive accuracy. Further work needs to be done to investigate turbulence modeling in PINNs so that high Reynolds number problems can be solved in a data-free manner. There are also many themes like uncertainty quantification \cite{UQ1, UQ2, UQ3} of surrogates and effective surrogate modeling of different geometries \cite{oldenburg2022geometry,kashefi2021point,gao2021phygeonet} that are active fields of research in PINNs, which could be included in future works that build on these results. \\

\noindent
\textbf{{\Large Acknowledgements}}\\
\linebreak
This research did not receive any specific grant from funding agencies in the public or not-for-profit sectors, or from any external commercial entities. The authors gratefully acknowledge the use of Altair Engineering Inc.'s computing facilities for running experiments.\\

\noindent
\textbf{{\Large CRediT authorship contribution statement}}\\
\linebreak

\textbf{Saakaar Bhatnagar:} Formal Analysis, Investigation, Methodology, Software, Validation, Writing-original draft. \textbf{Andrew Comerford:} Conceptualization, Investigation, Project Administration, Supervision, Writing- review and editing
\textbf{Araz Banaeizadeh:} Conceptualization, Project Administration, Supervision, Writing- review and editing
\bibliographystyle{unsrtnat}
\bibliography{bib}
\appendix

\section{Appendix}

\subsection{PINN model architecture and training details}
\label{sec:pinn_setup_training}
Every PINN model trained was created using fully connected layers. There were two sizes of models used:

\begin{enumerate}
    \item For models simulating a flow (as in Section \ref{sec:stenosis_problem},\ref{sec:pcb_solve} and \ref{sec:heat_sink_design}), a network 3 layers deep with 128 hidden units per layer was used.
    \item For models simulating the temperature in a body (as in \ref{sec:heat_sink_design}), a network 2 layers deep with 64 hidden units was used for each body.

\end{enumerate}

This distinction was made because
\begin{enumerate}
    \item Fluid domains tend to have more nodes due to being larger and having special mesh structures such as boundary layers.
    \item The solution in the flow domain tends to be more complex and hence needs more representation capacity.
    \item For thermal solves we use a domain decomposition strategy (described below) so the network representing each subdomain itself can be smaller, which leads to faster training.
\end{enumerate}
For more information on the multi-model domain decomposition approach, see Section \ref{sec:domain_decomp}.

The optimizer used was the Adam \cite{adam} optimizer, and the training points and data were divided into mini-batches of a randomly selected subset of the total training points, usually of size 1-5\% of the total available points. These mini batches were re-sampled every 5,000 training epochs. The initial learning rate was set to be \(10^{-3}\) with a decay factor of 0.9 after every 10,000 steps. Gradient clipping by global norm was also utilized to enhance the stability of the training process if large gradients occurred during the training. A warm start approach for training using the physics-based losses was used to reduce training time and cost (see Section \ref{subsec:warm_start})

\subsubsection{Warm Start for Physics}
\label{subsec:warm_start}
One of the issues with using standard initialization schemes like Xavier \cite{xavier_init} for training a PINN in a data-free manner, is that the resulting outputs of the networks have no physical meaning at the start of the training. Minimizing residual-based losses, in this case, is very ineffective since the gradients generated by backpropagation are likely to be very noisy and increase the likelihood of the training converging to a bad local minimum. One solution is to have a "warm start" by first training on some solution data only. This has the benefit of bringing network weights closer to a "converged value" before using the physics for training, which saves on computational cost

\begin{figure}[h!]
\centering
\begin{subfigure}{0.7\textwidth}
    \includegraphics[width=\textwidth]{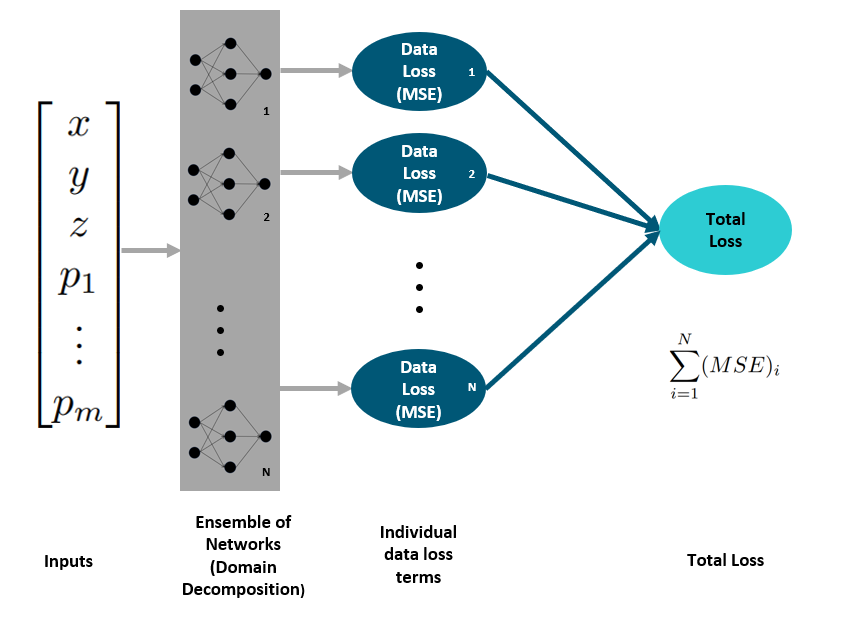}
    \caption{}
    \label{fig:training_data_graph}
\end{subfigure}
\hfill
\begin{subfigure}{0.85\textwidth}
    \includegraphics[width=\textwidth]{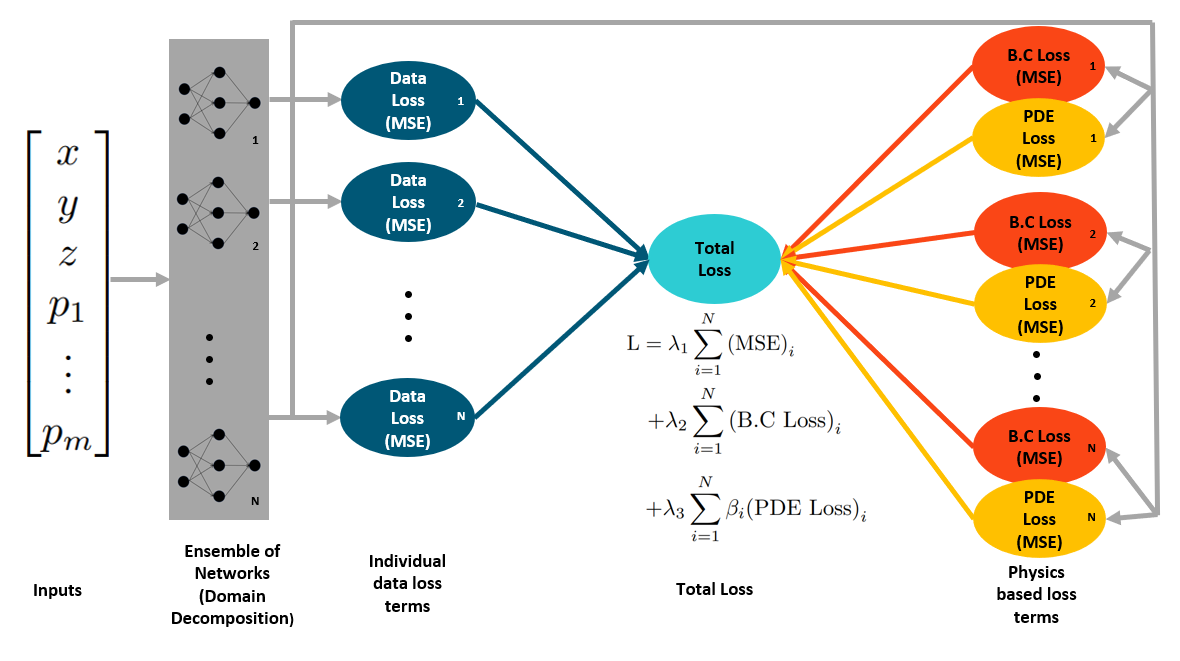}
    \caption{}
    \label{fig:training_data_physics_graph}
\end{subfigure}
\hfill
\caption{\centering Training graphs during different training phases. (\subref{fig:training_data_graph}) Training graph for warm start (\subref{fig:training_data_physics_graph}) Training graph when physics is included }
\label{fig:graphs}
\end{figure}

Figure \ref{fig:graphs} shows the training graphs for the warm starts period and the physics-informed learning period.

\subsubsection{Domain Decomposition}
\label{sec:domain_decomp}

The domain decomposition process is defined as breaking the overall solution domain into several subdomains and defining a model for each subdomain. As discussed by \citet{cPINN}and \citet{Hu_2022}, breaking down the overall solution domain into subdomains can allow for better generalization capability for complex multiphysics and multiscale problems since predictions for each subdomain are performed on the sub-network of that domain. This step is also crucial to enable the addition of interface boundary conditions to the loss function, which for the energy equation consists of continuity of flux and continuity of temperature.

Although there are several ways of implementing domain decomposition based on the problem being solved, in this work we focus on applying it to thermal surrogates only. This is so that we can apply the interface boundary conditions (Equations \ref{eqn:interface_1},\ref{eqn:interface_2}), and better capture the temperature field across different bodies that are a part of the same simulation, separated by interfaces.

\begin{figure}[h!]
\centering
\begin{subfigure}{0.9\textwidth}
    \includegraphics[width=\textwidth]{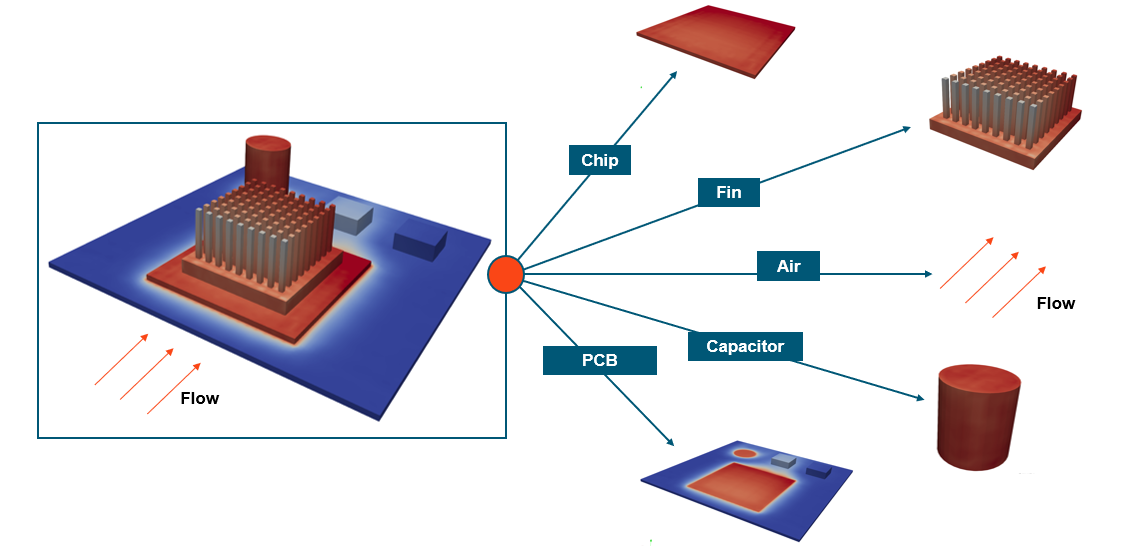}
    \caption{}
    
\end{subfigure}
\hfill

\caption{\centering Graphic showing the domain decomposition process for a Printed Circuit Board thermal surrogate modeling problem. Every part of the solution domain depicted has a different NN model predicting the solution field in it. }
\label{fig:domain_decomp_desc}
\end{figure}

\subsection{Design Optimizer Setup}

\label{subsec:PSO}
To optimize physical designs with trained models that predict flow-thermal results, the  Particle Swarm Optimization (PSO) \cite{PSO} algorithm is used. It is a zero-order optimization method that uses an initial distribution of particles in the search space and based on the "fitness" of each particle computes new positions and velocities of particles in the search space. Eventually, the particles converge towards the optima.

 The current categories of problems being solved in this paper (i.e constrained flow-thermal design optimization problems) take the general form

\[\min_{\textbf{u}} \; f(\textbf{u}),\]

s.t 

\[g_{i}(\textbf{u}) \leq X_{i} \;\;\;\;\; i=1.....N,\]

\[ u_{j}^{min} \leq u_{j} \leq u_{j}^{max} \;\;\;\; j=1..M,\]

where \(f\) represents an objective function, \(g_{i}\) represents the ith constraint and \(X_{i}\) represents constraint values. \textbf{u} represents the input vector of design parameters (of length M), and each component of \textbf{u} has a \(u_{j}^{min}\) and \(u_{j}^{max}\) that they can take. The objective and constraint values would be a derivative of flow thermal variables like pressure, temperature, or velocities, or design variables like geometry lengths, inflow rates, or source terms. The objective and constraints can be placed on bodies, individual surfaces, or the internal volumes of components that are a part of the simulation.

In order to solve this problem via the PSO algorithm, the constrained optimization problem is converted to an unconstrained problem via the penalty method to get the objective function:

\[f(\textbf{u}) + \sum_{i=1}^{N} \lambda_{i} \cdot (g_{i}(\textbf{u})-X_{i})^{2} \cdot \mathbb{1}(g_{i}(\textbf{u}) > X_{i}) + \beta \sum_{i=1}^{N} \mathbb{1}(g_{i}(\textbf{u}) > X_{i}) .\]

The first term is the constrained objective function. The second term represents the degree of deviation of the constraint from the boundary if the constraint is violated. The third term adds cost based on the number of constraints violated, and \(\lambda_{i}\) and \(\beta\) are constants.

Once the objective function is set up, the ith particle positions (\textbf{u}) and velocities are updated according to the equations:

\[\textbf{u}^{i}=\textbf{u}^{i}+\textbf{v}^{i},\]

\[\textbf{v}^{i}=w\textbf{v}^{i}+c_{1}r_{1}(\textbf{u}^{i}_{best}-\textbf{u}^{i})+ c_{2}r_{2}(\textbf{u}_{best}-\textbf{u}^{i}),\]

where \(c_{1}\), \(c_{2}\), \(r_{1}\), \(r_{2}\) and \(w\) are constants.\\

There are several reasons why PSO is chosen for design optimization:

\begin{enumerate}
    \item It is more economical to use compared to brute force grid search of Design of Experiment (DoE) space via querying solutions from the ANN. While this may not seem intuitive, in cases where there is pre-processing required (of say the point cloud) before the ANN can be queried for the result, reducing the number of model queries helps to minimize the amount of computation required. An example of this is when parametrizing geometry, in which case to query a new geometry, a new point cloud/mesh would have to be generated.
    
    \item  The distribution of particles at convergence gives a smaller subspace to do high-fidelity modeling, rather than returning a single particle as the best solution. This is important because the modeling process via ANNs is not exact or high fidelity, and it is better to have a subset of the DoE space returned rather than a single point, as for example, a gradient-based optimization method may do. Moreover, there may be multiple regions of the DoE space that satisfy the constraints and minimize the objective, and the PSO method can return both regions.

\end{enumerate}

\begin{figure}[h!]
  \centering
  \begin{minipage}[b]{0.6\textwidth}
    \includegraphics[width=\textwidth]{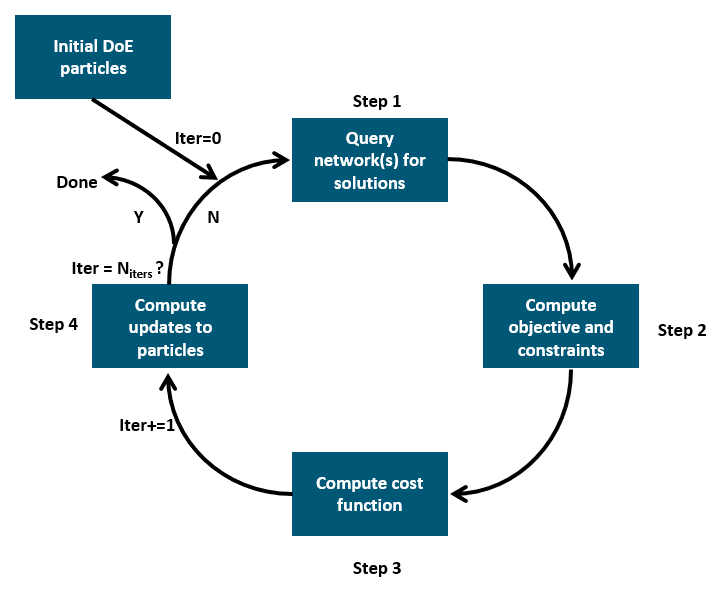}
    \caption{\centering A design iteration using a design optimization algorithm (like PSO)}
    \label{fig:PSO_cycle}
  \end{minipage}
  \hfill
  
\end{figure}

Figure \ref{fig:PSO_cycle} depicts the design cycle using the PSO algorithm. In traditional design optimization, Step 1 is done using CFD solvers, but this can get very expensive. NN-based surrogates are an ideal tool to replace CFD solvers for early-stage design cycles.

\subsection{Extended: Surrogate Modeling and Design Optimization of a Heat Sink via PINNs}

\subsubsection{Design of Experiments (DoE) tables}
\label{sec:doe_tables}

Tables \ref{table:surrogate_training_table} and \ref{table:surrogate_testing_table} show the parameters used to train and query the model. The data for the training points was generated by running Altair AcuSolve\textsuperscript{\textregistered}. Each solution took roughly 15 minutes on a 4-core machine.

\begin{table}[h!]
\begin{center}
\begin{tabular}{ |c|c|c|c| } 
 \hline
  No. & Inflow Velocity (m/s)  & Fin Height (mm)  & Power Generated (W)  \\ 
 \hline
  1 & 5 & 19 &  45 \\ 
  \hline
  2 & 3 & 15 &  30 \\
  \hline
  3 & 3 & 15 & 60 \\
  \hline
  4 & 3 & 23 & 30 \\
  \hline
  5 & 3 & 23 & 60 \\
  \hline
  6 & 7 & 15 & 30 \\
  \hline
  7 & 7 & 15 & 60 \\
  \hline
  8 & 7 & 23 & 30 \\
  \hline
  9 & 7 & 23 & 60 \\
  \hline
  10 & 5.03 & 15.3 & 59 \\
  \hline
  11 & 3.01 & 18.4 & 45.55 \\
  \hline
  12 & 6.94 & 19.4 & 36.05 \\
  \hline
  13 & 6.53 & 22.6 & 51 \\
 \hline
\end{tabular}
\caption{DoE table to train the surrogate model.}
\label{table:surrogate_training_table}
\end{center}
\end{table}

\begin{table}[h!]
\begin{center}
\begin{tabular}{ |c|c|c|c| } 
 \hline

 No. & Inflow Velocity (m/s)  & Fin Height (mm)  & Power Generated (W)  \\
 
 \hline
  Test Point 1 & 6 & 20 & 47.5 \\ 
  \hline
  Test Point 2 & 4 & 17 &  40 \\
  \hline
  Test Point 3 & 6.5 & 22 & 55 \\
 \hline
  Test Point 4 & 3.5 & 19 & 50 \\
 \hline
\end{tabular}
\caption{Test points to query the model.}
\label{table:surrogate_testing_table}
\end{center}
\end{table}

\subsubsection{Cost Benefit Analysis of Surrogate Model Compared to CFD Based Optimization}
\label{sec:cost_benefit_analysis_heat_sink}

 A quantitative comparison of PINN surrogate versus CFD-based optimization is shown. For the PINNs model training was performed on a single Titan V GPU card.

\begin{figure}[h!]
  \centering
  \begin{minipage}[b]{0.5\textwidth}
    \includegraphics[width=\textwidth]{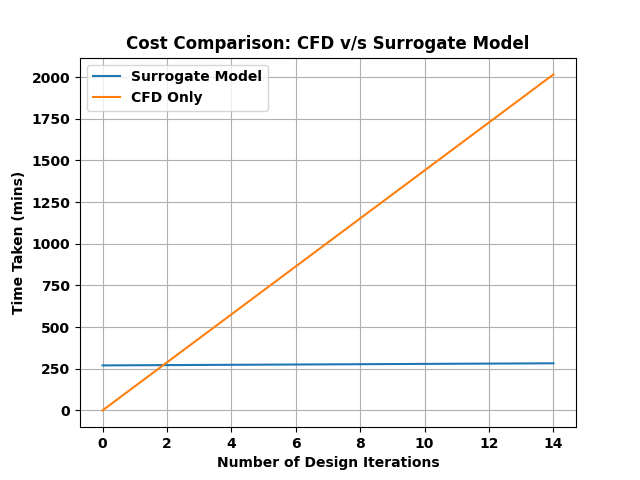}
    
  \end{minipage}
  \caption{\centering Time comparison of doing PSO-based design iterations using the surrogate versus CFD. Once the surrogate returns a truncated DoE space, the designer can perform high-fidelity CFD to fine-tune the design.}
  \hfill
  \label{fig:heatsink_costs_comparison}
  
\end{figure}

Table \ref{table:heatsink_costs_comparison} shows the cost-benefit analysis of the hybrid PINNs-CFD surrogate model versus CFD for the heat sink design optimization problem shown in Section ~\ref{sec:heat_sink_design}. The comparison, visualized in Figure \ref{fig:heatsink_costs_comparison} shows the total time taken to optimize the design using the PSO method and compares the time taken versus the number of iterations. The "model training time" encapsulates the time taken to create the data and train the model, so that it is a fair comparison.

\begin{table}[h!]
\begin{center}
\begin{tabular}{ |c|c|c|c| } 
 \hline
 Solve Type  & Model Training Time  & \makecell{Time for a \\ Design Iteration} & \makecell{Time for 10 \\ Design Iterations}   \\ 
 \hline
  CFD (4 cores) & - & 160 min & 1600 min \\ 
  \hline
  PINN (1 GPU + 1 core) & 286 mins & 55 seconds &  300 mins \\
  \hline
  
\end{tabular}
\caption{Comparing design iteration times using CFD versus ANN surrogate for the heatsink problem.}
\label{table:heatsink_costs_comparison}
\end{center}
\end{table}

\subsection{Improvements Above Data Driven NNs}
\label{sec:imp_above_baseline}

 Some additional qualitative comparisons for accuracy comparisons are shown. NN models are created with the core features described in Section \ref{sec:impfeatures} and compared to baseline data-driven NNs.
 
\begin{figure}[h!]
\centering
\begin{subfigure}[t]{0.45\textwidth}
    \includegraphics[width=\textwidth]{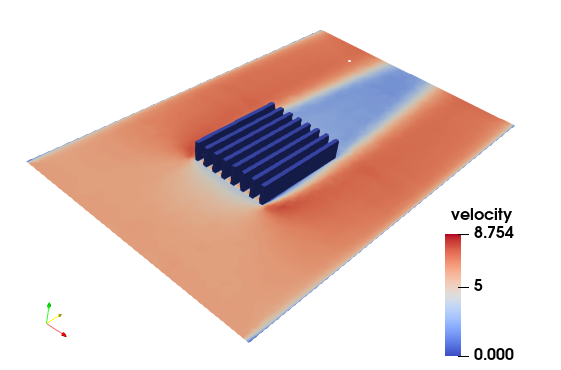}
    \caption{\centering }
    \label{fig:sink_vel_PINN}
\end{subfigure}
\hfill
\begin{subfigure}[t]{0.45\textwidth}
    \includegraphics[width=\textwidth]{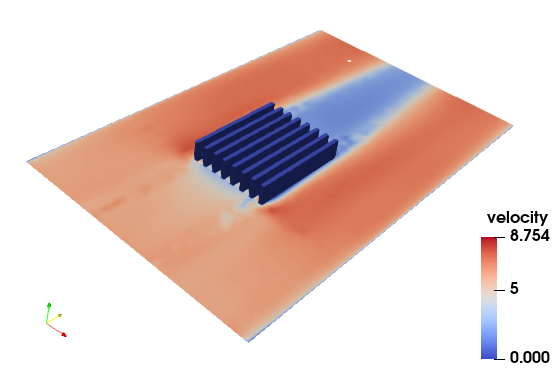}
    \caption{\centering }
    \label{fig:sink_vel_normal}
\end{subfigure}
\hfill
\begin{subfigure}[t]{0.45\textwidth}
    \includegraphics[width=\textwidth]{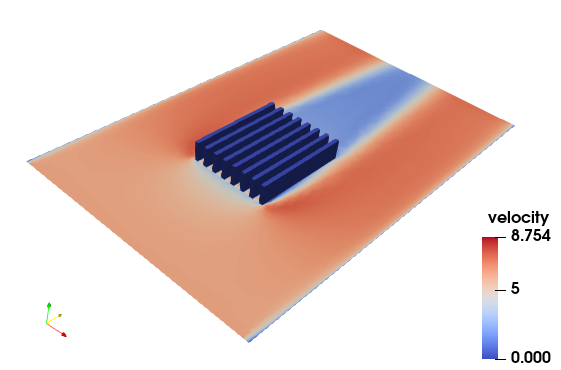}
    \caption{\centering  }
    \label{fig:sink_vel_true}
\end{subfigure}
\hfill
\begin{subfigure}[t]{0.45\textwidth}
    \centering
    \includegraphics[width=0.6\textwidth]{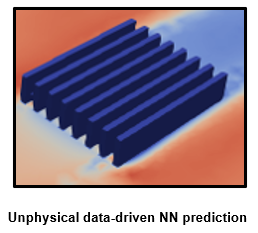}
    \caption{\centering }
    \label{fig:sink_vel_inset}
\end{subfigure}
\caption{\centering Velocity prediction from standard data-driven NN versus PINN-based networks for Test Point 1 (Table \ref{table:surrogate_test_pt_errors}). (\subref{fig:pcb_temp_PINN}) Prediction from PINN 
(\subref{fig:pcb_temp_normal}) Prediction from standard fully connected NN 
(\subref{fig:pcb_temp_true}) True Solution
(\subref{fig:pcb_temp_inset}) Zoom-in from the prediction of standard NN showing unphysical results, especially near the boundaries. }
\label{fig:sink_vel_accuracy}
\end{figure}

\begin{figure}[h!]
\centering
\begin{subfigure}[t]{0.45\textwidth}
    \includegraphics[width=\textwidth]{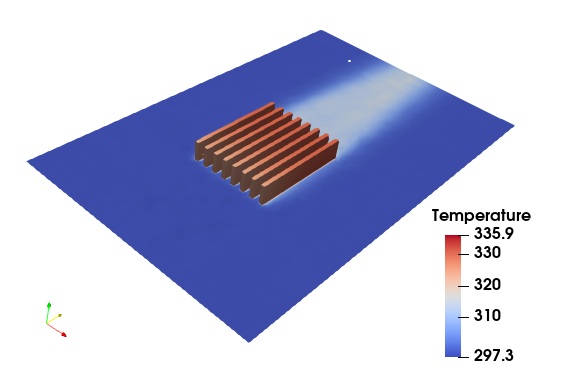}
    \caption{\centering }
    \label{fig:sink_temp_PINN}
\end{subfigure}
\hfill
\begin{subfigure}[t]{0.45\textwidth}
    \includegraphics[width=\textwidth]{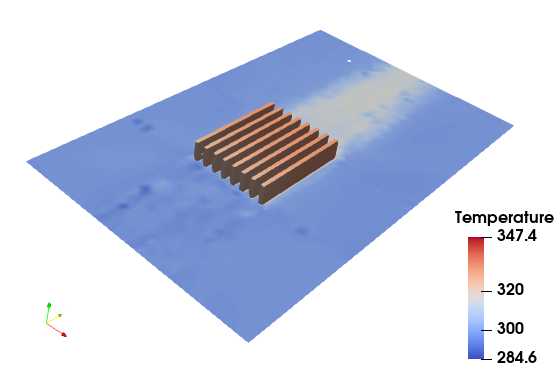}
    \caption{\centering }
    \label{fig:sink_temp_normal}
\end{subfigure}
\hfill
\begin{subfigure}[t]{0.45\textwidth}
    \includegraphics[width=\textwidth]{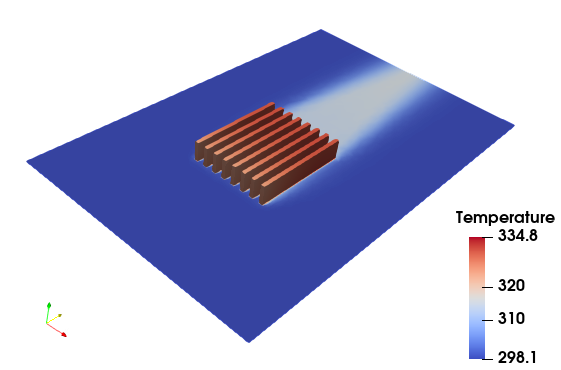}
    \caption{\centering  }
    \label{fig:sink_temp_true}
\end{subfigure}
\hfill
\begin{subfigure}[t]{0.45\textwidth}
    \includegraphics[width=\textwidth]{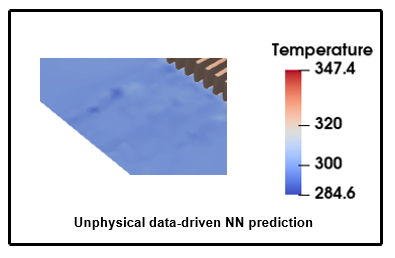}
    \caption{\centering }
    \label{fig:sink_temp_inset}
\end{subfigure}
\caption{\centering Temperature prediction on the heat sink geometry from standard data-driven NN and PINN. (\subref{fig:pcb_temp_PINN}) Prediction from PINN 
(\subref{fig:pcb_temp_normal}) Prediction from standard fully connected NN 
(\subref{fig:pcb_temp_true}) True Solution
(\subref{fig:pcb_temp_inset}) Zoom-in from the prediction of standard NN showing unphysical results, especially near the boundaries. }
\label{fig:sink_temp_accuracy}
\end{figure}

\begin{figure}[h!]
\centering
\begin{subfigure}[t]{0.45\textwidth}
    \includegraphics[width=\textwidth]{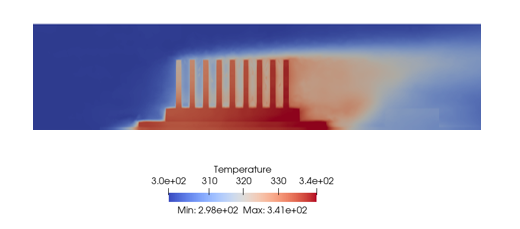}
    \caption{\centering }
    \label{fig:pcb_temp_PINN}
\end{subfigure}
\hfill
\begin{subfigure}[t]{0.45\textwidth}
    \includegraphics[width=\textwidth]{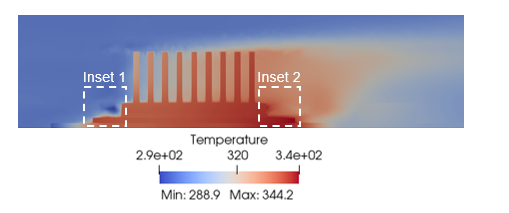}
    \caption{\centering }
    \label{fig:pcb_temp_normal}
\end{subfigure}
\hfill
\begin{subfigure}[t]{0.45\textwidth}
    \includegraphics[width=\textwidth]{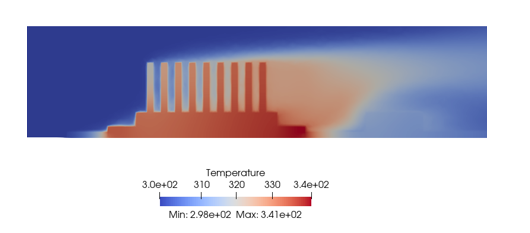}
    \caption{\centering  }
    \label{fig:pcb_temp_true}
\end{subfigure}
\hfill
\begin{subfigure}[t]{0.45\textwidth}
    \includegraphics[width=\textwidth]{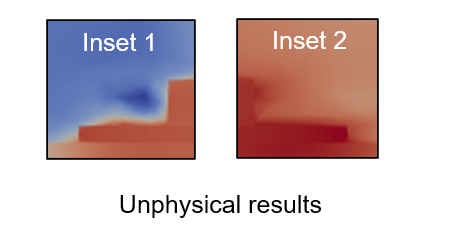}
    \caption{\centering }
    \label{fig:pcb_temp_inset}
\end{subfigure}
\caption{\centering Temperature comparison on the PCB between predictions from standard data-driven NN versus PINN-based networks. (\subref{fig:pcb_temp_PINN}) Prediction from PINN 
(\subref{fig:pcb_temp_normal}) Prediction from standard fully connected NN 
(\subref{fig:pcb_temp_true}) True Solution
(\subref{fig:pcb_temp_inset}) Insets from the prediction of standard NN showing unphysical results, especially near the boundaries }
\label{fig:temp_pcb_accuracy}
\end{figure}

Figures \ref{fig:sink_vel_accuracy}, \ref{fig:sink_temp_accuracy} and  \ref{fig:temp_pcb_accuracy} show example qualitative comparisons of PINN models versus standard data-driven models. The results from the standard data-driven model have regions of non-physical results.

\end{document}